\documentclass[12pt]{iopart}
\usepackage[dvips]{graphicx}
\usepackage{iopams,setstack}

\begin{document}
\title[The one-loop effective action in $\phi^4$ theory coupled non-linearly 
with curvature]{
The one-loop effective action  in $\phi^4$ theory coupled non-linearly 
with curvature power and dynamical origin of cosmological constant}
\author{
T Inagaki\dag\footnote[4]{Also at Hiroshima Univ., Higashi-Hiroshima, Japan},
S Nojiri\ddag  and S D Odintsov\S\footnote[5]{Also at TSPU, Tomsk, Russia}
}
\address{\dag\ School of Chemistry, Physics and Earth Science, Flinders 
University, Adelaide, SA 5001, AUSTRALIA}
\address{\ddag\ Department of Applied Physics, National Defence Academy,
Hashirimizu Yokosuka 239-8686, JAPAN}
\address{\P\ Instituci\`o Catalana de Recerca i Estudis Avan\c{c}ats 
and Institut d'Estudis Espacials de Catalunya, 
Edifici Nexus, Gran Capit\`a 2-4, 08034 Barcelona, SPAIN}

\eads{\mailto{inagaki@hiroshima-u.ac.jp},
\mailto{snojiri@yukawa.kyoto-u.ac.jp},
\mailto{odintsov@aliga.ieec.uab.es}}

\begin{abstract}
The functional formulation and one-loop effective action
for scalar self-interacting theory non-linearly coupled with some power
of the curvature are studied. After the
explicit one-loop renormalization at weak curvature,
we investigated numerically the phase structure for such unusual $\phi^4$
theory. It is demonstrated the possibility of curvature-induced phase
transitions for positive values of the curvature power,
while for negative values the radiative symmetry breaking does not take
place. The dynamical mechanism for the explanation of the current
smallness of the cosmological constant is presented for several
models from the class of theories under consideration.
\end{abstract}


\maketitle

\section{Introduction}
The current stage of theoretical cosmology and, in particularly, the dark energy
problem calls to consideration of new field theory models
which maybe useful for the explanation of the  universe acceleration. In the
situation when the attempts to use the
well-known classes of field theories as candidates for dark energy (dark
matter) are not quite successful, such strategy seems to be justified.
Moreover, one should bear in mind that number of problems of the early time
cosmology (say, the fundamental cosmological constant problem)
are still too far from the resolution. The (old or new) field theoretical model
which should include also gravitation itself should solve these problems
at once
(at least, in the regime where one can still neglect the
quantum gravity effects).

In the recent paper \cite{NO} the class of dark energy models where dark energy
(scalar or spinor field theory, etc) couples with some power of the curvature
has been proposed. It has been demonstrated on the example of scalar Lagrangian
coupled with the power of curvature that such modification
of gravity-matter coupling term may explain not only the current cosmic
speed-up, but also the dark energy dominance assisted by gravity. On the same
time, theories of similar nature have been recently proposed as good candidate
for dynamical resolution of the cosmological constant problem \cite{dolgov,MR}
(see also \cite{J}). Of course, such models are not standard ones,
in the sense that they are not multiplicatively renormalizable in curved
spacetime \cite{BOS}. Hence, they should be considered as kind of effective
theories (without clear understanding of their origin and their relation with
more fundamental string/M-theory).

In the present paper we continue the study of the properties of such
models on the example of scalar self-interacting theory non-linearly
coupled with some power of the curvature. Admitting the discrete symmetry
which prohibits odd powers of scalar in the potential, the effective
action formulation (section 2) is developed in close analogy with
multiplicatively
renormalizable theories \cite{BOS}. Using Riemann normal coordinate
expansion, the modified Green function is evaluated at weak curvature
limit (section 3). (We work in linear curvature approximation what is
expected to be appropriate for the study of quantum effects in the
inflationary universe and in late-time, dark energy universe).
Subsequently, this Green function is used for the calculation of
the one-loop effective action in the same limit (section 4). After the one-loop
renormalization of the effective action done in section 5, one arrives at
finite expression
which is used for the numerical study of the phase structure in FRW
universe. The number of examples explicitly demonstrates the behaviour of
the effective potential for such effective theory. In particularly, it is
shown the possibility of curvature-induced phase transitions in the model
under discussion (section 6). It is indicated that radiative corrections are
not very
relevant at weak curvature.

Finally, in section 7 it is shown that scalar theory under consideration may
serve also
for dynamical resolution of cosmological constant problem. The explicit
choice of parameters (which maybe considered as kind of fine-tuning)
giving such dynamical resolution is presented. It is interesting that
radiative corrections are not relative for such a mechanism.
Some discussion and outlook are given in the last section.

\section{$\phi^{4}$ theory coupled non-linearly with curvature power}

We extend the $\phi^{4}$ theory which is the simplest model where
the spontaneous symmetry breaking takes place.
One of the possible couplings of scalar Lagrangian with curvature (at some
power) maybe included into the starting action:
\begin{equation}
    S = \int d^D x \ \sqrt{-g} \left[\frac{1}{2\kappa^2}R
        + \left(\frac{R}{M^2}\right)^\alpha {\cal L}_d \right] .
\label{a0}
\end{equation}
where $M$ is an arbitrary mass scale, $g$ is the determinant of the metric
tensor $g_{\mu\nu}$ and ${\cal L}_d$ is the ordinary Lagrangian density of
the $\phi^{4}$ theory,
\begin{equation}
    {\cal L}_d(\phi_0) = \frac{1}{2}\phi_{0;\mu}{\phi_0}^{;\mu}
    - \frac{\xi_0 R}{2} \phi_0^2
    +\frac{\mu_0^2}{2}\phi_0^2
    -\frac{\lambda_0}{4!}\phi_0^4,
\label{I2}
\end{equation}
where $\phi_0$ is a real scalar field.
Our sign convention for the metric is $(+---\cdots)$ and follows the notation
of Birrell and Davies \cite{BD}.
If the spacetime curvature is constant, the non-linear curvature coupling
disappears by the transformation
\begin{eqnarray}
     \phi & \rightarrow & \left(\frac{R}{M^2}\right)^{-\alpha/2} \phi,
\\
     \lambda_0 & \rightarrow & \left(\frac{R}{M^2}\right)^{\alpha} \lambda_0 .
\end{eqnarray}
We are interested in a non-trivial effect of the non-linear curvature coupling.
It will be found in a non-static and/or inhomogeneous spacetime.
Note that the non-linear coupling of above sort has been proposed
in Ref. \cite{NO} (see also related model in \cite{ANO}) as the model of dark
energy which naturally
resolves the problem of dark energy dominance in the current universe.

The Lagrangian is invariant under the discrete transformation,
\begin{equation}
    \phi \rightarrow -\phi .
\end{equation}
This symmetry prevents the Lagrangian from having $\phi^{3}$ terms.
A non-vanishing expectation value for the field $\phi$ breaks
the discrete symmetry spontaneously. To evaluate the expectation
value for $\phi$ we calculate the effective action.

We start from the generating functional $W[J]$.
The generating functional of the theory is given by
\begin{equation}
    e^{\frac{i}{\hbar}W[J]}
    \equiv \int {\cal D} \phi_0 \
    e^{\frac{i}{\hbar}S+\frac{i}{\hbar}\int d^{D}x \sqrt{-g} \phi_0 J}
\label{GeF}
\end{equation}
In the presence of the source $J$ the classical equation of motion becomes
\begin{equation}
    \left.\frac{\delta S}{\delta \phi_0(x)}\right|_{\phi_0  = \phi_b} = -J(x).
\label{EoM}
\end{equation}
In curved spacetime the functional derivative is defined by
\begin{equation}
    S[\phi_0+\delta\phi] - S[\phi_0]
    = \int d^{D}x \sqrt{-g} \frac{\delta S}{\delta \phi_0(x)} \delta \phi_0(x).
\end{equation}

We divide the field $\phi_0$ into a classical background
$\phi_{b}$ which satisfies the Eq.(\ref{EoM})
and a quantum fluctuation $\tilde{\phi}$,
\begin{equation}
\phi_0 = \phi_{b} + \hbar^{1/2}\tilde{\phi}.
\end{equation}
In terms of $\phi_{b}$ and $\tilde{\phi}$
the action $S[\phi_0]$ is rewritten as
\begin{eqnarray}
    S[\phi_0] &=& S[\phi_{b} + \hbar^{1/2}\tilde{\phi}]
\nonumber \\
    &=& S[\phi_{b}]
      +\int d^{D}x \sqrt{-g} \left.\frac{\delta S[\phi_0]}{\delta
\phi_0(x)}\right|_{\phi_0  = \phi_b}\hbar^{1/2}\tilde{\phi(x)}
\nonumber \\
    &&+\frac{\hbar}{2}\int d^{D}x \int d^{D}y \sqrt{-g(x)} \sqrt{-g(y)}
      \tilde{\phi}(x)iG_d^{-1}(x,y)\tilde{\phi}(y)
     +\mbox{O}({\tilde{\phi}^3})
\nonumber \\
    &=& S[\phi_{b}]
      -\int d^{D}x \sqrt{-g}J(x) \hbar^{1/2}\tilde{\phi}(x)
\nonumber \\
    &&+\frac{\hbar}{2}\int d^{D}x \int d^{D}y \sqrt{-g(x)} \sqrt{-g(y)}
      \tilde{\phi}(x)iG_d^{-1}(x,y)\tilde{\phi}(y)
\nonumber \\
    && +\mbox{O}({\hbar^{3/2}\tilde{\phi}^3}),
\end{eqnarray}
where $G_d^{-1}(x,y)$ is given by
\begin{eqnarray}
    iG_d^{-1}(x,y) & \equiv &
    \frac{\delta^2 S[\phi_0]}{\delta \phi_0(x)\delta\phi_0(y)}
\nonumber \\
    & = & -\sqrt{-g(x)}\left[
          \left(\frac{R}{M^2}\right)^\alpha
          \left(\square + \xi_0 R - \mu_0^2
          +\frac{\lambda_0}{2}\phi_b^2 \right)\right.
\nonumber \\
    &  & + \left. \alpha\left(\frac{R}{M^2}\right)^{\alpha-1}
          \frac{R_{;\mu}}{M^2}\nabla^{\mu}\right].
\label{gd}
\end{eqnarray}
Therefore the generating functional (\ref{GeF}) is expanded to be
\begin{eqnarray}
    e^{\frac{i}{\hbar}W[J]} &=& e^{\frac{i}{\hbar}\left(S[\phi_{b}]
      -\int d^{D}x \sqrt{-g}J(x) \phi_b(x) \right)}
\nonumber \\
    && \times \int {\cal D} \tilde{\phi}\ e^{\frac{i}{2}
      \int d^{D}x \int d^{D}y \sqrt{-g(x)} \sqrt{-g(y)}
      \tilde{\phi}(x)iG_d^{-1}(x,y)\tilde{\phi}(y)+\mbox{O}(\hbar^{1/2})}
\end{eqnarray}
Performing the path-integral,
\begin{equation}
    \int {\cal D} \tilde{\phi}\ e^{\frac{i}{2}
    \int d^{D}x \int d^{D}y \sqrt{-g(x)}\sqrt{-g(y)}
    \tilde{\phi}(x)iG_d^{-1}(x,y)\tilde{\phi}(y)}
    = \left[\mbox{Det}\ iG_d^{-1}\right]^{-1/2} ,
\end{equation}
we obtain the generating functional
\begin{equation}
    W[J] = S[\phi_{b}]
    -\int d^{D}x \sqrt{-g}J(x) \phi_b(x)
    -\frac{i\hbar}{2}\ln \left[\mbox{Det}\ iG_d^{-1}\right]
    +\mbox{O}(\hbar^{3/2}).
\label{gf0}
\end{equation}

The effective action $\Gamma [\phi_c]$ is given by the Legendre transform
of $W[J]$,
\begin{equation}
    \Gamma [\phi_c] \equiv W[J] - \int d^{D}x \sqrt{-g} \phi_c(x)J(x),
\label{ea0}
\end{equation}
where $\phi_c$ denotes the expectation value of $\phi_0$ in the
presence of the source $J$,
\begin{equation}
    \phi_c \equiv \frac{\delta W[J]}{\delta J}.
\end{equation}
Substituting Eq. (\ref{gf0}) to Eq. (\ref{ea0}) we find that the effective
action is expanded to be
\begin{equation}
    \Gamma [\phi_c] = S[\phi_c]
    -\frac{i\hbar}{2}\ln \left[\mbox{Det}\ iG_d^{-1}\right]
    +\mbox{O}(\hbar^{3/2}).
\label{ea}
\end{equation}

In the path integral formalism the expectation value of $\phi_0$ in the
ground state satisfies
\begin{equation}
    \langle \phi_0 \rangle \equiv \lim_{J\rightarrow 0} \frac{\delta
W[J]}{\delta
J}.
\end{equation}
This equation is rewritten as
\begin{equation}
    \left. \frac{\delta \Gamma[\phi_c]}{\delta \phi_c} \right|_{\phi_c
\rightarrow \langle \phi_0 \rangle} = 0 ,
\label{Gap:Action}
\end{equation}
where we have used the relation
\begin{equation}
    \frac{\delta \Gamma[\phi_c]}{\delta \phi_c} = J.
\end{equation}
The equation (\ref{Gap:Action}) is called the gap equation.
The solution of the gap equation shows the presence of stationary points
of the effective action $\Gamma[\phi_c]$. Thus, the phase
structure of the theory maybe found by observing the stationary points of
$\Gamma[\phi_c]$ (for a review of the quantum field theory in curved 
spacetime, see Ref. \cite{BOS}).

\section{Scalar Green function at the weak curvature limit}

Following the procedure developed in Refs. \cite{BP, PT}, we solve the
Klein-Gordon equation and calculate the Green function $G_d$ (\ref{gd})
at the weak curvature limit. Note that in Green function calculation we
actually
work in the expansion on the curvatures what is known to be appropriate
for the study of quantum effects in the inflationary universe and also
in late-time, dark energy universe.

The Green function $G_d$ satisfies the modified Klein-Gordon equation,
\begin{eqnarray}
    & &   \sqrt{-g(x)}\left[
          \left(\frac{R}{M^2}\right)^\alpha
          \left(\square + \xi_0 R - \mu_0^2
          +\frac{\lambda_0}{2}\phi^2 \right)\right.
\nonumber \\
    &  & \hspace*{2cm} + \left. \alpha\left(\frac{R}{M^2}\right)^{\alpha-1}
          \frac{R_{;\mu}}{M^2}\nabla^{\mu}\right]iG_d(x,x')=\delta^D(x-x').
\label{KG}
\end{eqnarray}
For a practical calculations it is more convenient to introduce $G(x,x')$
and $H_x$ which are defined by
\begin{equation}
    G(x,x')\equiv (-g)^{1/4}(x)\left(\frac{R(x)}{M^2}\right)^{\alpha/2}
                  iG_d(x,x')
                  (-g)^{1/4}(x')\left(\frac{R(x')}{M^2}\right)^{\alpha/2} ,
\label{gre00}
\end{equation}
and
\begin{eqnarray}
    H_x &=& (-g)^{1/4}(x)\left(\frac{R(x)}{M^2}\right)^{\alpha/2}
          \sqrt{-g(x)}\left[
          \left(\square + \xi_0 R - \mu_0^2
          +\frac{\lambda_0}{2}\phi^2 \right)\right.
\nonumber \\
    &  &  \hspace*{18mm} + \left. \alpha\left(\frac{R}{M^2}\right)^{-1}
          \frac{R_{;\mu}}{M^2}\nabla^{\mu}\right]
          (-g)^{-1/4}(x)\left(\frac{R(x)}{M^2}\right)^{-\alpha/2} ,
\label{ham00}
\end{eqnarray}
In terms of $G(x,x')$ and $H_x$ the Klein-Gordon equation (\ref{KG}) reads
\begin{equation}
    H_x G(x,x') = \delta^D (x-x') .
\label{KG2}
\end{equation}

After some calculations the Hamiltonian density $H_x$ is rewritten as
\begin{equation}
    H_x = \partial_\mu g^{\mu\nu} \partial_\nu
         +\alpha\frac{R_{;\mu}}{R}\partial^{\mu}+ V_x ,
\end{equation}
where the potential $V_x$ is given by
\begin{eqnarray}
    V_x &=&
\left[(-g)^{1/4}\partial_\mu(-g)^{1/2}g^{\mu\nu}\partial_\nu(-g)^{-1/4}
          \right]
          + \xi_0 R - \mu_0^2 +\frac{\lambda_0}{2}\phi^2
    \nonumber \\
      &&  + \alpha \frac{R_{;\mu}}{R} \left[(-g)^{-1/4}\partial^\mu(-g)^{1/4}
                                      \right]
          -\frac{\alpha}{2}\left(\frac{\alpha}{2}-1\right)
          \frac{R^{;\mu}R_{;\mu}}{R^2}
          -\frac{\alpha}{2}\frac{\square R}{R} .
\end{eqnarray}

Here we adopt the Riemann normal coordinate expansion \cite{P}.
We set the origin of the coordinates at $x$ and write
$y=x-x'$ and neglect terms proportional to $y^a$, $a\geq 2$.
Hence, the Eq.(\ref{KG2}) is rewritten as
\begin{equation}
    \left(\eta_{\mu\nu}\partial^\mu\partial^\nu + \zeta_\mu\partial^\mu
    + \xi_{\mu\nu} y^\mu \partial^\nu
    +\chi^2  +\beta_\mu y^\mu \right) G(y) = \delta^D (y) ,
\label{KG3}
\end{equation}
where
\begin{eqnarray}
    \zeta_\mu &=& \left.\alpha \frac{R_{;\mu}}{R}\right|_{y=0} ,
\\
    \xi_{\mu\nu} &=& \left.\alpha \left( \frac{R_{;\mu\nu}}{R}
                   -\frac{R_{;\mu}R_{;\nu}}{R^2} \right) \right|_{y=0},
\\
    \chi^2 & = & \left.\left[ - \mu_0^2 +\frac{\lambda_0}{2}\phi^2
             + \left(\xi_0-\frac{1}{6}\right) R
             -\frac{\alpha}{2}\left(\frac{\alpha}{2}-1\right)
              \frac{R^{;\mu}R_{;\mu}}{R^2}
             -\frac{\alpha}{2}\frac{\square R}{R} \right]\right|_{y=0}
\nonumber \\
          & \equiv & -\mu_0^2 + U ,
\\
    \beta_\mu &=& \left. \left(
    U_{;\mu} + \frac{\alpha}{6}\frac{R^{;\nu}}{R}R_{\mu\nu}
    \right)\right|_{y=0}.
\end{eqnarray}

Performing the Fourier transformation
\begin{equation}
    G(y)=\int \frac{d^D p}{(2\pi)^D} e^{i p y} \tilde{G}(p),
\end{equation}
one finds that the equation (\ref{KG3}) becomes
\begin{equation}
    \left(-p^2 + i\zeta_\mu p^\mu  + \xi_{\mu\nu}  p^\nu \partial_p^\mu
    + \chi^2  -i \beta_\mu \partial_p^\mu \right) \tilde{G}(p) = 1 .
\label{KG4}
\end{equation}

First we consider the solution for a positive $\chi^2$. In this case
   the following proper-time form for the solution of
Eq.(\ref{KG4}) is assumed
\begin{equation}
    \tilde{G}(p) = \int^{\infty}_{0} ds e^{-\chi^2 s}
                   e^{p^{\mu}A_{\mu\nu}(s)p^{\nu}+i B_{\mu}(s)p^{\mu}+C(s)} .
\label{ass}
\end{equation}
Substituting this  into the Eq.(\ref{KG4}) one obtains
\begin{eqnarray}
    && \int^{\infty}_{0} ds \left(
         -p^2 + i\zeta_\mu p^\mu  + \xi_{\mu\nu} X^{\mu} p^\nu
         + \chi^2  -i \beta_\mu X^{\mu}
       \right)
\nonumber \\
    && \hspace*{10mm} \times
       e^{-\chi^2 s}e^{p^{\mu}A_{\mu\nu}(s)p^{\nu}+i B_{\mu}(s)p^{\mu}+C(s)}
\nonumber \\
    && = 1 ,
\label{as2}
\end{eqnarray}
where
\begin{equation}
    X^\mu = A^{\mu\nu} p_\nu + p_\nu A^{\mu\nu} + i B^{\mu} .
\end{equation}
On the other hand, there is an identity
\begin{eqnarray}
    && - \int^{\infty}_{0} ds \frac{d}{ds}
       e^{-\chi^2 s}e^{p^{\mu}A_{\mu\nu}(s)p^{\nu}+i B_{\mu}(s)p^{\mu}+C(s)}
\nonumber \\
    && = \int^{\infty}_{0} ds \left( \chi^2
                              - p^\mu \frac{dA_{\mu\nu}}{ds} p^\nu
                              - i \frac{dB_{\mu}}{ds} p^\mu
                              - \frac{dC}{ds} \right)
\nonumber \\
    && \hspace*{10mm} \times
       e^{-\chi^2 s}e^{p^{\mu}A_{\mu\nu}(s)p^{\nu}+i B_{\mu}(s)p^{\mu}+C(s)}
\nonumber \\
    && = 1 ,
\label{as3}
\end{eqnarray}
Comparing the Eq.(\ref{as2}) with the Eq.(\ref{as3}),
the simultaneous differential equations are obtained
\begin{eqnarray}
    \frac{dA_{\mu\nu}}{ds} & = & 1 - \xi_{\mu\rho}
                               ( {A^{\rho}}_{\nu} + {A_{\nu}}^{\rho}) ,
\\
    \frac{dB_{\mu}}{ds} & = & -\zeta_{\mu} - \xi_{\mu\rho} B^\rho
                              -\beta_{\rho}
                              ( {A^{\rho}}_{\nu} + {A_{\nu}}^{\rho}) ,
\\
\frac{dC}{ds} & = & - \beta_{\mu} B^{\mu}.
\end{eqnarray}
The solutions of these differential equations are found to be
\begin{eqnarray}
    A & = & \frac{1}{2} \xi^{-1}+C_1 e^{-2\xi s} ,
\\
    B & = & -\xi^{-1}\zeta + \beta \xi^{-2} +C_2 e^{-\xi s}
            -2\beta\xi^{-1}C_1 e^{-2\xi s} ,
\\
    C & = & \beta\xi^{-1}\zeta s -(\beta\xi^{-1})^2 s
            +\beta\xi^{-1}C_2 e^{-\xi s}
            - (\beta\xi^{-1})^2 C_1 e^{-2\xi s}
\nonumber \\
    &&  +C_3 ,
\end{eqnarray}
where $C_1$, $C_2$ and $C_3$ are constant matrix parameters which are
determined by the boundary conditions.
In the flat space time, $R \rightarrow 0$, Eq.(\ref{KG4}) must
reproduce a Klein-Gordon equation in the Minkowski space.
Thus, the functions $A$, $B$ and $C$ must satisfy
\begin{eqnarray}
A & \rightarrow & s ,
\\
B & \rightarrow & 0 ,
\\
C & \rightarrow & 0 ,
\label{bc}
\end{eqnarray}
at the limit $\zeta$, $\xi$, $\beta \rightarrow 0$.

These boundary conditions fix the parameters $C_1$, $C_2$
and $C_3$ to be
\begin{eqnarray}
    C_1 & = & -\frac{1}{2}\xi^{-1} ,
\\
    C_2 & = & -2\beta\xi^{-2} + \xi^{-1}\zeta ,
\\
    C_3 & = & \frac{3}{2} (\beta\xi^{-1})(\beta\xi^{-2}) - \beta\xi^{-2}\zeta .
\end{eqnarray}
Therefore the functions $A$, $B$ and $C$ are simplified to
\begin{eqnarray}
    A & = & \frac{1}{2} \xi^{-1} (1- e^{-2\xi s}) ,
\\
    B & = & -\xi^{-1}\zeta (1 - e^{-\xi s})
            +\beta\xi^{-2} (1- e^{-\xi s})^2 ,
\\
    C & = & -\beta\xi^{-2}\zeta (1 - e^{-\xi s} - \xi s)
\nonumber \\
    &&      +2 (\beta\xi^{-1})(\beta\xi^{-2}) \left[
            \left(1 - e^{-\xi s} - \xi s\right)  - \frac{1}{4}\left(
            1 - e^{-2\xi s} -2 \xi s
            \right)\right].
\end{eqnarray}

Since the parameters $\zeta$, $\xi$ and $\beta$ are small enough
at the weak curvature limit, we neglect the higher order term about
these parameters for a small $s$ and get
\begin{eqnarray}
    A & \sim & s - \xi s^2 ,
\\
    B & \sim & -\zeta s + \beta s^2 ,
\\
    C & \sim & \frac{1}{2}\beta\zeta s^2 - \frac{1}{3}\beta^2 s^3 \sim 0 .
\end{eqnarray}
Inserting these solutions into Eq.(\ref{ass}),
   the Green function $\tilde{G}(p)$ follows
\begin{equation}
    \tilde{G}(p) = \int^{\Lambda}_{0}ds
    e^{-(p^{\mu}\xi_{\mu\nu}p^{\nu}-i\beta_\mu p^\mu)s^2
       +(p^2-\chi^2-i\zeta_\mu p^\mu)s } ,
\label{gpp}
\end{equation}
where a scale $\Lambda\sim \sqrt{|R|/M^2}$ is introduced. It corresponds to
the upper limit where we can neglect the higher order term about the
parameters $\zeta$, $\xi$ and $\beta$. Since the integrand becomes smaller
and smaller above the scale $\Lambda$, we drop the contribution from the region
above
$\Lambda$.
It is straightforward to extend above analysis to a negative $\chi^2$ and
obtain
\begin{equation}
    \tilde{G}(p) = \int^{-\Lambda}_{0}ds
    e^{-(p^{\mu}\xi_{\mu\nu}p^{\nu}-i\beta_\mu p^\mu)s^2
       +(p^2-\chi^2-i\zeta_\mu p^\mu)s } ,
\label{gpp2}
\end{equation}

Integrating over $s$ in Eq.(\ref{gpp}) and (\ref{gpp2}), it follows
\begin{eqnarray}
    \tilde{G}(p) &=&
    \sqrt{\frac{\pi}{4(p^{\mu}\xi_{\mu\nu}p^{\nu}-i\beta_\mu p^\mu)}}
    \exp\left[\frac{(p^2-\chi^2-i\zeta_\mu
p^\mu)^2}{4(p^{\mu}\xi_{\mu\nu}p^{\nu}-i\beta_\mu p^\mu)}\right]
\nonumber \\
    &&\times
    \left[1+\mbox{erf}\left(-\mbox{\scriptsize sgn}(\chi^2)\frac{p^2-\chi^2-i\zeta_\mu
p^\mu}
      {\sqrt{4(p^{\mu}\xi_{\mu\nu}p^{\nu}-i\beta_\mu p^\mu)}}\right)\right] ,
\end{eqnarray}
where erf$(z)$ is the error function.
In the configuration space the Green function $G(y)$ reads
\begin{equation}
    G(y) = \int^{\mbox{\scriptsize sgn}(\chi^2)\Lambda}_{0}ds e^{-\chi^2 s }\int \frac{d^D
p}{(2\pi)^D}
    e^{ip^\mu y_\mu}e^{p^\mu A_{\mu\nu} p^\nu +i B_\mu p^\mu} .
\end{equation}
At the origin it reduces to
\begin{eqnarray}
    G(0) & = & \int^{\mbox{\scriptsize sgn}(\chi^2)\Lambda}_{0}ds e^{-\chi^2 s }
               \int \frac{d^D p}{(2\pi)^D}
               e^{p^\mu A_{\mu\nu} p^\nu +i B_\mu p^\mu}
\nonumber \\
         & = & \frac{i}{(2\pi)^D} \int^{\mbox{\scriptsize sgn}(\chi^2)\Lambda}_{0} ds
               \left(\det \frac{A}{\pi}\right)^{-1/2}
               e^{-\frac{1}{4}BA^{-1}B-\chi^2 s} .
\end{eqnarray}
To perform the momentum integral  the Wick rotation maybe applied.
Neglecting the higher order term about $\zeta$, $\xi$ and $\beta$,
we obtain
\begin{eqnarray}
    G(0) \sim \frac{i}{(4\pi)^{D/2}} \int^{\mbox{\scriptsize sgn}(\chi^2)\Lambda}_{0}
              \frac{ds}{s^{D/2}}
              \left( 1+\frac{1}{2} \mbox{tr} \xi s \right)
               e^{-\chi^2 s} .
\label{g0}
\end{eqnarray}
Hence, the scalar Green function is found at small curvature. It will be used
in the calculation of the one-loop effective action.

\section{The one-loop effective action}

The effective action of the present model is given by (\ref{ea}).
Here we normalize the effective action to satisfy
\begin{equation}
    \Gamma[\phi = 0] = S[\phi = 0].
\end{equation}
In this normalization the one-loop effective action is given by
\begin{equation}
    \Gamma [\phi] = S[\phi]
    -\frac{i\hbar}{2}\left[\ln \left\{\mbox{Det}\ iG_d^{-1}(\phi^2)\right\}
    -\ln \left\{\mbox{Det}\ iG_d^{-1}(\phi^2 = 0)\right\}\right] .
\label{ea2}
\end{equation}
The right hand side of Eq. (\ref{ea2}) contains a term proportional to
\begin{eqnarray}
    && \ln \left\{\mbox{Det}\ iG_d^{-1}(\phi^2) \right\}
    - \ln \left\{\mbox{Det}\ iG_d^{-1}(\phi^2 = 0) \right\}
\nonumber \\
   && = \mbox{Tr} \left\{\ln  iG_d^{-1}(\phi^2) \right\}
    - \mbox{Tr} \left\{\ln  iG_d^{-1}(\phi^2 = 0) \right\}
\nonumber \\
    && = -\frac{\lambda_0}{2} \int d^D x (-g)
         \left(\frac{R}{M^2}\right)^{\alpha}
         \int^{\phi^2}_{0} dm^2\
         iG_d(x,x;\phi^2 = m^2) .
\end{eqnarray}

From the Eq.(\ref{g0}) the Green function at the weak curvature limit
is found to be
\begin{equation}
    G(x,x;\phi^2 = m^2) = 
     \frac{i}{(4\pi)^{D/2}}
     \int^{\mbox{\scriptsize sgn}(\chi^2)\Lambda}_{0}
              \frac{ds}{s^{D/2}}
              \left( 1+\frac{1}{2} \mbox{tr} \xi s \right)
               e^{-\chi^2(\phi^2 = m^2) s} .
\end{equation}
The relationship between $G_d$ and $G$ is given by
\begin{equation}
    iG_d(x,x;\phi^2 = m^2) =
     \frac{1}{\sqrt{-g(x)}}\left(\frac{R(x)}{M^2}\right)^{-\alpha}
     G(x,x;\phi^2 = m^2) .
\label{gd2}
\end{equation}
The effective action at the weak curvature limit reduces to
\begin{eqnarray}
    \Gamma [\phi] & = & S[\phi]
    -\frac{\hbar\lambda_0}{4(4\pi)^{D/2}}
\\
\nonumber
    &&  \times
        \int d^D x \sqrt{-g}\int^{\phi^2}_{0} dm^2
        \int^{\mbox{\scriptsize sgn}(\chi^2)\Lambda}_{0}
              \frac{ds}{s^{D/2}}
              \left( 1+\frac{1}{2} \mbox{tr} \xi s \right)
               e^{-\chi^2(\phi^2=m^2) s} .
\label{ea3}
\end{eqnarray}
Integrating over $m^2$ one gets
\begin{eqnarray}
    \Gamma [\phi] & = & S[\phi]
    +\frac{\hbar}{2(4\pi)^{D/2}}
    \int d^D x \sqrt{-g}
\nonumber \\
    &&  \times
        \int^{\mbox{\scriptsize sgn}(\chi^2)\Lambda}_{0}
              \frac{ds}{s^{D/2+1}}
              \left( 1+\frac{1}{2} \mbox{tr} \xi s \right)
               \left(e^{-\chi^2(\phi^2) s}-e^{-\chi^2(\phi^2=0) s}
               \right) .
\label{ea4p}
\end{eqnarray}
Performing the integration about $s$ and neglecting the higher
order term about $1/\Lambda$, we get
\begin{eqnarray}
    \Gamma [\phi] & = & S[\phi]
\nonumber \\
   && +\frac{\hbar}{2(4\pi)^{D/2}}\Gamma\left(-\frac{D}{2}\right)
    \int d^D x \sqrt{-g}\left[
       (\chi^2(\phi^2))^{D/2}
              \left( 1-\frac{D \mbox{tr} \xi}{4\chi^2(\phi^2)}\right)\right.
\nonumber \\
   && \left. \hspace*{2cm}
       - (\chi^2(\phi^2=0))^{D/2}
              \left( 1-\frac{D \mbox{tr} \xi}{4\chi^2(\phi^2=0)}\right) \right]
.
\label{ea4}
\end{eqnarray}

The effective Lagrangian density is defined by
\begin{equation}
    \Gamma[\phi]=\int d^D x \sqrt{-g} {\cal L}_{eff}.
\end{equation}
From Eqs.(\ref{a0}) and (\ref{ea4}) it is found to be
\begin{eqnarray}
    {\cal L}_{eff} &=&
        \frac{1}{2\kappa^2}R
        + \left(\frac{R}{M^2}\right)^\alpha \left(
        \frac{1}{2}\phi_{0;\mu}{\phi_0}^{;\mu}
        - \frac{\xi_0 R}{2} \phi_0^2
        +\frac{\mu_0^2}{2}\phi_0^2
        -\frac{\lambda_0}{4!}\phi_0^4\right)
\nonumber \\
    &&  +\frac{\hbar}{2(4\pi)^{D/2}}\Gamma\left(-\frac{D}{2}\right)
       \left[(\chi^2(\phi^2))^{D/2}
              \left( 1-\frac{D \mbox{tr} \xi}{4\chi^2(\phi^2)} \right) \right.
\nonumber \\
    &&  \left. - (\chi^2(\phi^2=0))^{D/2}
              \left( 1-\frac{D \mbox{tr} \xi}{4\chi^2(\phi^2=0)} \right)\right] .
\label{al}
\end{eqnarray}
with
\begin{eqnarray}
    \mbox{tr} \xi &=& \alpha \left( \frac{\square R}{R}
                   -\frac{R^{;\mu}R_{;\mu}}{R^2} \right) ,
\\
    \chi^2(\phi^2) & = & - \mu_0^2 +\frac{\lambda_0}{2}\phi^2
             + \left(\xi_0-\frac{1}{6}\right) R
             -\frac{\alpha}{2}\left(\frac{\alpha}{2}-1\right)
              \frac{R^{;\mu}R_{;\mu}}{R^2}
             -\frac{\alpha}{2}\frac{\square R}{R} ,
\end{eqnarray}
In the limit $\alpha\rightarrow 0$ and $R\rightarrow 0$ the effective
Lagrangian density (\ref{al}) reproduces the one in the Minkowski
space.

\section{The one-loop renormalization}
The effective action (\ref{ea4}) and the effective
Lagrangian density (\ref{al}) are divergent in two and
four dimensions. To obtain the finite value we must renormalize the
theory. (Remind that theory is not multiplicatively renormalizable in curved
spacetime \cite{BOS}). At the present order we impose the following
renormalization conditions
\begin{equation}
    \left. \frac{\partial^2 \Gamma[\phi]}{\partial \phi^2} \right|_{\phi=M_1}
    \equiv \left(\frac{R}{M^2}\right)^\alpha (\mu_r^2 - \xi_r R) ,
\label{def:mur}
\end{equation}
\begin{equation}
    \left. \frac{\partial^4 \Gamma[\phi]}{\partial \phi^4} \right|_{\phi=M_2}
    \equiv -\left(\frac{R}{M^2}\right)^\alpha \lambda_r .
\label{def:lr}
\end{equation}
where $M_1$ and $M_2$ are the renormalization scales.%
\footnote{
For a higher order calculation we must impose the other conditions to
renormalize the wave function for the field $\phi$ and the cosmological
constant.
}
From these conditions
one obtains the relationship between the renormalized parameters
$\mu_r$, $\xi_r$, $\lambda_r$ and the bare ones
$\mu_0$, $\xi_0$, $\lambda_0$.
\begin{eqnarray}
\mu_r^2 - \xi_r R &=& \mu_0^2 - \xi_0 R -\frac{\lambda_0}{2}M_1^2
    -\frac{\hbar}{2(4\pi)^{D/2}}\left(\frac{R}{M^2}\right)^{-\alpha}
    \lambda_0
\nonumber \\
   && \times \left\{\Gamma\left(1-\frac{D}{2}\right)
    (\chi^2(M_1^2))^{D/2-1}
\right.
\nonumber \\
   && \left.\hspace*{18mm}\times
\left[
    1+\lambda_0 M_1^2\left(\frac{D}{2}-1\right)(\chi^2(M_1^2))^{-1}\right]
    \right.
\nonumber\\
   && \left. +\frac{1}{2}\Gamma\left(2-\frac{D}{2}\right)
    (\chi^2(M_1^2))^{D/2-2}
\right.
\nonumber \\
   && \left.\hspace*{18mm}\times
\left[
    1+\lambda_0 M_1^2\left(\frac{D}{2}-2\right)(\chi^2(M_1^2))^{-1}\right]
    \mbox{tr}\xi
    \right\}
\nonumber \\
&\equiv& \mu_0^2 - \xi_0 R -\frac{\lambda_0}{2}M_1^2
    +\frac{\hbar}{2(4\pi)^{D/2}}\left(\frac{R}{M^2}\right)^{-\alpha}
    \lambda_0 f(D,M_1),
\end{eqnarray}
\begin{eqnarray}
\lambda_r &=& \lambda_0
    -\frac{\hbar}{2(4\pi)^{D/2}}\left(\frac{R}{M^2}\right)^{-\alpha}
    \lambda_0^2
\nonumber \\
   && \times \left\{\Gamma\left(2-\frac{D}{2}\right)
    (\chi^2(M_2^2))^{D/2-2}\left[
    3+6\lambda_0 M_2^2\left(\frac{D}{2}-2\right)(\chi^2(M_2^2))^{-1}
\right.\right.
\nonumber \\
   && \left.\left.\hspace*{28mm}
    +\lambda_0^2 M_2^4\left(\frac{D}{2}-2\right)\left(\frac{D}{2}-3\right)
    (\chi^2(M_2^2))^{-2}
    \right]
    \right.
\nonumber\\
   && \left. +\frac{1}{2}\Gamma\left(3-\frac{D}{2}\right)
    (\chi^2(M_2^2))^{D/2-3}\left[
    3+6\lambda_0 M_2^2\left(\frac{D}{2}-3\right)(\chi^2(M_2^2))^{-1}
\right.\right.
\nonumber \\
   && \left.\left.\hspace*{28mm}
    +\lambda_0^2 M_2^4\left(\frac{D}{2}-3\right)\left(\frac{D}{2}-4\right)
    (\chi^2(M_2^2))^{-2}
    \right]
    \mbox{tr}\xi
    \right\}
\nonumber \\
&\equiv& \lambda_0
    +\frac{\hbar}{2(4\pi)^{D/2}}\left(\frac{R}{M^2}\right)^{-\alpha}
    \lambda_0^2 g(D,M_2) .
\end{eqnarray}

By using these renormalized parameters the effective Lagrangian density
(\ref{al}) is rewritten as
\begin{eqnarray}
    {\cal L}_{eff} &=&
        \frac{1}{2\kappa^2}R
        + \left(\frac{R}{M^2}\right)^\alpha \left(
        \frac{1}{2}\phi_{;\mu}{\phi}^{;\mu}
        - \frac{\xi_r R}{2} \phi^2
        +\frac{\mu_r^2}{2}\phi^2
        +\frac{\lambda_r}{4}M_1^2 \phi^2
        -\frac{\lambda_r}{4!}\phi^4\right)
\nonumber \\
    &&  -\frac{\hbar}{4(4\pi)^{D/2}}\lambda_r f(D,M_1) \phi^2
        +\frac{\hbar}{48(4\pi)^{D/2}}\lambda_r^2 g(D,M_2) (\phi^4 -6M_1^2\phi^2)
\nonumber \\
    &&  +\frac{\hbar}{2(4\pi)^{D/2}}\Gamma\left(-\frac{D}{2}\right)
       \left[(\chi^2(\phi^2))^{D/2}
              \left( 1-\frac{D \mbox{tr} \xi}{4\chi^2(\phi^2)} \right) \right.
\nonumber \\
    &&  \left. - (\chi^2(0))^{D/2}
              \left( 1-\frac{D \mbox{tr} \xi}{4\chi^2(0)} \right)\right] .
\label{alr}
\end{eqnarray}
At the two-dimensional limit Eq.(\ref{alr}) becomes
\begin{eqnarray}
    {\cal L}_{eff}^{2D}
         &=&
        \frac{1}{2\kappa^2}R
        + \left(\frac{R}{M^2}\right)^\alpha \left(
        \frac{1}{2}\phi_{;\mu}{\phi}^{;\mu}
        - \frac{\xi_r R}{2} \phi^2
        +\frac{\mu_r^2}{2}\phi^2
        +\frac{\lambda_r}{4} M_1^2 \phi^2
        -\frac{\lambda_r}{4!}\phi^4\right)
\nonumber \\
    &&  +\frac{\hbar}{8\pi}\left[
        -\frac{1}{2}\lambda_r \phi^2
        \left(1-\ln\frac{\chi^2(\phi^2)}{\chi^2(M_1^2)}\right)
         +\left(\chi^2(0)-\frac{1}{2}\mbox{tr}\xi\right)
         \ln \frac{\chi^2(\phi^2)}{\chi^2(0)}
        \right.
\nonumber \\
    &&  -\frac{1}{2}\lambda_r^2 \phi^2 M_1^2\frac{1}{\chi^2(M_1^2)}
        +\frac{1}{4}\lambda_r \phi^2 \frac{1}{\chi^2(M_1^2)}
         \left(1-\lambda_r M_1^2\frac{1}{\chi^2(M_1^2)}\right)\mbox{tr}\xi
\nonumber \\
    &&  -\frac{1}{8}\lambda_r^2 (\phi^4-6M_1^2\phi^2) \frac{1}{\chi^2(M_2^2)}
\label{alrd2}
\\
    &&  \times\left\{\frac{1}{3}
        \left(3-6\lambda_r M_2^2\frac{1}{\chi^2(M_2^2)}
              +2\lambda_r^2 M_2^4\frac{1}{(\chi^2(M_2^2))^2}
        \right)\right.
\nonumber \\
\nonumber
    &&  \left. \left.+\frac{1}{2\chi^2(M_2^2)}
        \left(1-4\lambda_r M_2^2\frac{1}{\chi^2(M_2^2)}
              +2\lambda_r^2 M_2^4\frac{1}{(\chi^2(M_2^2))^2}\right)
        \mbox{tr}\xi \right\} \right] .
\end{eqnarray}
Taking the four-dimensional limit of Eq.(\ref{alr}), we obtain
   \begin{eqnarray}
    {\cal L}_{eff}^{4D} &=&
        \frac{1}{2\kappa^2}R
        + \left(\frac{R}{M^2}\right)^\alpha \left(
        \frac{1}{2}\phi_{;\mu}{\phi}^{;\mu}
        - \frac{\xi_r R}{2} \phi^2
        +\frac{\mu_r^2}{2}\phi^2
        +\frac{\lambda_r}{4}M_1^2 \phi^2
        -\frac{\lambda_r}{4!}\phi^4\right)
\nonumber \\
    &&  +\frac{\hbar}{128\pi^2}\left[
        \lambda_r^2 (\phi^4-6M_1^2\phi^2) \left(
        \frac{\lambda_r M_2^2}{\chi^2(M_2^2)}
        -\frac{1}{6}\frac{\lambda_r^2 M_2^4}{\chi^2(M_2^2)}
        \right)
        \right.
\nonumber \\
    &&  +3 \lambda_r\phi^2 \chi^2(0)
        +\frac{3}{4}\lambda_r^2\phi^4
        -2\lambda_r \phi^2 \chi^2(M_1^2)
        -2(\chi^2(0))^2\ln\frac{\chi^2(\phi^2)}{\chi^2(0)}
\nonumber \\
    &&  -2\lambda_r\phi^2\chi^2(0) \ln \frac{\chi^2(\phi^2)}{\chi^2(M_1^2)}
        -\frac{1}{2}\lambda_r^2\phi^4 \ln \frac{\chi^2(\phi^2)}{\chi^2(M_2^2)}
\label{alrd4} \\
    &&  +3\lambda_r^2M_1^2\phi^2 \ln \frac{\chi^2(M_1^2)}{\chi^2(M_2^2)}
\nonumber \\
    &&  -\left\{
        \frac{\lambda_r^2}{4}(\phi^4-6M_1^2\phi^2) \left(
        \frac{1}{\chi^2(M_2^2)}
        -\frac{2\lambda_r M_2^2}{(\chi^2(M_2^2))^2}
        +\frac{2}{3}\frac{\lambda_r^2 M_2^4}{(\chi^2(M_2^2))^3}
        \right)\right.
\nonumber \\
    &&  \left.\left. +\lambda_r\phi^2
        +\frac{\lambda_r^2  M_1^2\phi^2}{\chi^2(M_1^2)}
        -2\chi^2(0)\ln\frac{\chi^2(\phi^2)}{\chi^2(0)}
        -\lambda_r\phi^2\ln\frac{\chi^2(\phi^2)}{\chi^2(M_1^2)}\right\}
        \mbox{tr}\xi \right] .
\nonumber
\end{eqnarray}

Therefore the ultra-violet divergences in the effective Lagrangian
density (\ref{al}) and the effective action (\ref{ea4}) disappear after the
one-loop renormalization. Note that to make sense to divergent expressions one
can apply any other regularization (say, zeta-regularization \cite{EORBZ}). Of
course, the question of dependence from the regularization in such effective
models remains.

\section{Phase Structure in FRW spacetime}

It is expected that the non-linear curvature coupling in our model leads to
non-trivial consequences for the phase structure in a non-static spacetime.
(This may have the interesting applications for early-time, accelerating
 and late-time, accelerating universe when quantum gravity effects are
negligible).
Here we consider the model in the $D$-dimensional Fridmann-Robertson-Walker
spacetime. By spirit it is similar to the study of phase structure
for scalar self-interacting theory, scalar electrodynamics and GUTs
in curved spacetime (see \cite{S, A, KI, HO, BO1, BO2, BO3} and for complete 
list of refs. and review
\cite{BOS}) where curvature-induced phase transitions were discovered.
The FRW spacetime is defined by the metric
\begin{equation}
    ds^2 = dt^2 - a^2(t)
    \left[\frac{dr^2}{1-K r^2}+r^2 d \Omega_{D-2}
    \right] ,
\end{equation}
where $K$ corresponds to the curvature of the $D-1$-dimensional space.
The curvature $R$ in this spacetime is given by
\begin{equation}
    R = 2(D-1)
        \left(\dot{H}+\frac{D}{2}H^2
              +\frac{D-2}{2}\frac{K}{a^2}\right) ,
\end{equation}
where the Hubble rate $H$ is defined by $H\equiv \dot a/a$.
The first derivative of the curvature is
\begin{eqnarray}
    R_{;0} &=& 2(D-1)\left[\ddot{H}+D\dot{H}H
               -(D-2)H\frac{K}{a^2}\right] ,
\\
    R_{;i} &=& 0 .
\end{eqnarray}
Thus, it follows
\begin{eqnarray}
    R^{;\mu}R_{;\mu} &=& 4(D-1)^2 \left[\ddot{H}+D\dot{H}H
               -(D-2)H\frac{K}{a^2}\right]^2 ,
\label{RR1}
\\
    \square R &=& 2(D-1) \left[\dddot{H}
    +(2D-1)\ddot{H}H
    +D(D-1)\dot{H}H^2+D\dot{H}^2
    \right.
\nonumber \\
    && \left.
    -(D-2)\left( D H^2+\dot{H}-3\frac{H}{a}\right)\frac{K}{a^2}\right] .
\label{RR2}
\end{eqnarray}

Here  the spacetime with $K=0$ and $a(t)=a_0 t^{h_0}$ is considered. In this
spacetime the curvature invariants are
\begin{eqnarray}
    R &=& 2(D-1)h_0\left(\frac{D}{2}h_0-1\right)\frac{1}{t^2},
\label{r:frw}
\\
    \frac{R^{;\mu}R_{;\mu}}{R^2} &=& 4 \frac{1}{t^2},
\\
    \frac{\square R}{R} &=& -2\left[(D-1)h_0 -3\right]
    \frac{1}{t^2}.
\end{eqnarray}

The ground state of the theory depends on the spacetime structure and the
parameters of the theory under consideration.
To study the phase structure of the $\phi^4$ theory in FRW spacetime we
evaluate the effective action, ${\cal L}_{eff}$.
The expectation value of the field $\phi$ is obtained by searching of the
stationary points of the effective action.
The effective action develops a small imaginary part for a negative
$\chi(\phi)$. We evaluate a real part of the effective action to find the
ground state.

In an expanding universe the scale factor $a(t)$ increases as the time runs.
Here we suppose that the scale factor $a(t)$ is proportional to the time $t$,
i.e. $a(t)=at$ and numerically calculate the renormalized effective action
(\ref{alr}). All  mass scales are normalized by an arbitrary mass scale $M$ and
$\hbar =1$. The renormalization scale is chosen to be $M_1 =0$ and $M_2 =0.1
M^{(D-2)/2}$.

First we consider the stationary and spatially homogeneous $\phi$.
In this case the kinetic term of $\phi$ disappears. One can define the ordinary
effective potential $V(\phi)$ by
\begin{equation}
     V(\phi)\equiv -{\cal L}_{eff}.
\end{equation}
The expectation value of the field $\phi$ is defined by the minimum of the
effective potential.

\begin{figure}[t]
\begin{center}
\begin{minipage}{6.8cm}
\includegraphics[width=\linewidth]{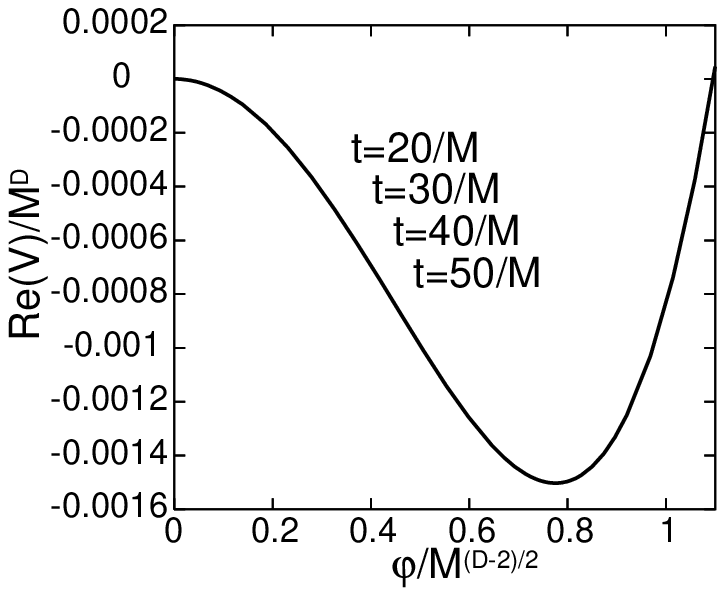}
\end{minipage}
\begin{minipage}{6.8cm}
\includegraphics[width=\linewidth]{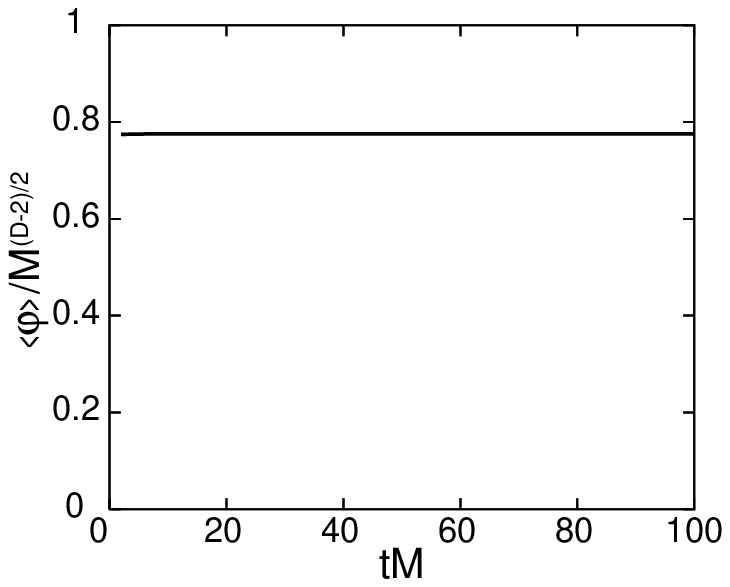}
\end{minipage}
\caption{\label{minimu01a0c1}Behaviour of the effective potential and the mass
gap for $\alpha=0$, $h_0=1$, $\xi=0$, $\mu_r=0.1 M$, $\lambda=0.1 M^{4-D}$ and
$D=3.8$.}
\end{center}
\end{figure}

\begin{figure}[t]
\begin{center}
\begin{minipage}{6.8cm}
\includegraphics[width=\linewidth]{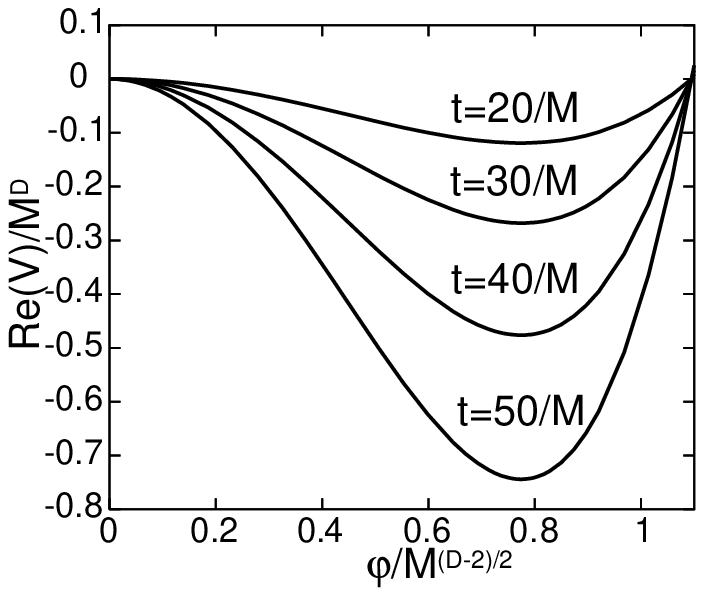}
\end{minipage}
\begin{minipage}{6.8cm}
\includegraphics[width=\linewidth]{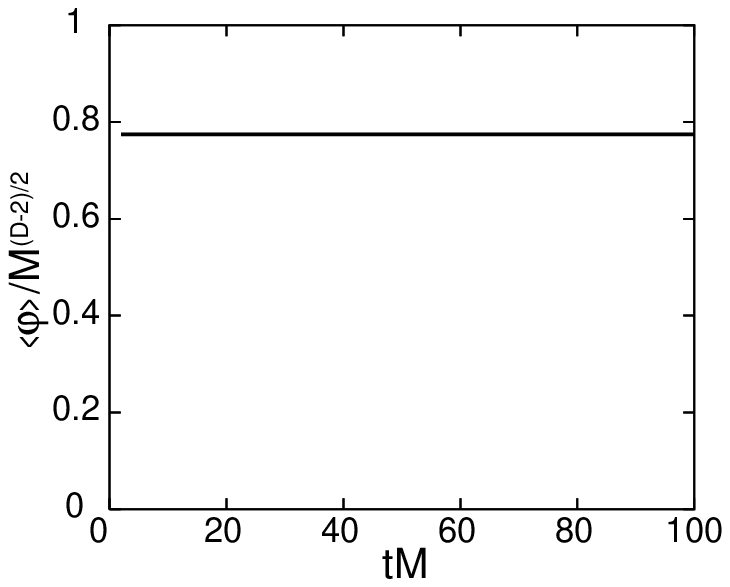}
\end{minipage}
\caption{\label{minimu01a-1c0}Behaviour of the effective potential and the mass
gap for $\alpha=-1$, $h_0=1$, $\xi=0$, $\mu_r=0.1 M$, $\lambda=0.1 M^{4-D}$ and
$D=3.8$.}
\end{center}
\end{figure}

\begin{figure}[t]
\begin{center}
\begin{minipage}{6.8cm}
\includegraphics[width=\linewidth]{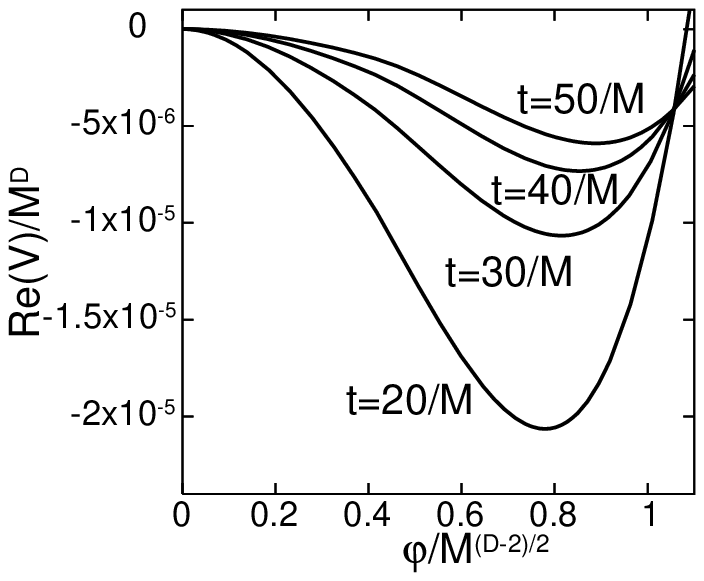}
\end{minipage}
\begin{minipage}{6.8cm}
\includegraphics[width=\linewidth]{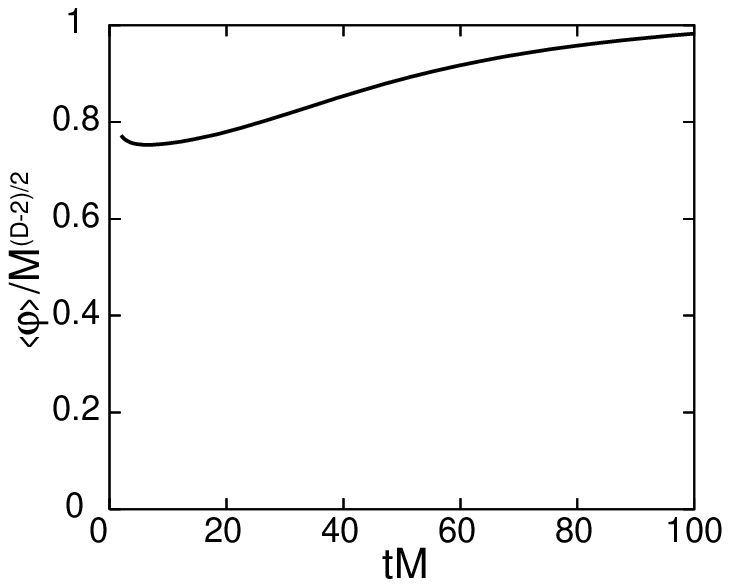}
\end{minipage}
\caption{\label{minimu01a1c1}Behaviour of the effective potential and the mass
gap for $\alpha=1$, $h_0=1$, $\xi=0$, $\mu_r=0.1 M$, $\lambda=0.1 M^{4-D}$ and
$D=3.8$.}
\end{center}
\end{figure}

We illustrate the behavior of the effective potential for a minimal
gravitational coupling $\xi=0$ at $\mu_r=0.1 M$ in Figs.1, 2 and 3. 
Here we set $D=3.8$ to regularize the theory with keeping the general
covariance. It corresponds to the cut-off scale, $\Lambda \sim 500 M_2$, 
where $M_2$ is a renormalization scale.
Since the spacetime curvature behaves as $t^{-2}$, it raises up for a 
small $t$.
It is not valid to apply the weak curvature expansion at the beginning of
the universe. In Figs. 1, 2 and 3 we solve the gap equation only for $2<tM$.
As is shown in the figures the effective potential seems to be almost
proportional to $R^\alpha \propto t^{-2\alpha}$.

In Figs. 1 and 2 the expectation value $\langle\phi\rangle$ is located near
the stationary point of the classical potential,
$\displaystyle \phi^2 = \frac{6}{\lambda}\mu_r^2$.
For the minimal gravitational coupling the classical part of the effective
action has no curvature dependence except for the overall factor
and the Einstein-Hilbert term. On the other side, the radiative correction
depends on the curvature.
At $\alpha =0$ our model reduces to the ordinary $\phi^4$ theory.
As is known, the radiative correction has only a small contribution to
spontaneous symmetry breaking for a large $\mu_r$ in the ordinary theory.
In the case of a negative $\alpha$ the classical part of the effective
action (\ref{alr}) is enhanced as the curvature decreases.
Contributions from the radiative corrections are lost at a large $t$.
For a positive $\alpha$ the classical part of the effective
action is suppressed except for the Einstein-Hilbert term as the curvature
decreases. Hence we observe the curvature dependence
of the expectation value $\langle\phi\rangle$ only if $\alpha >0$, see Fig. 3.

\begin{figure}[t]
\begin{center}
\begin{minipage}{6.8cm}
\includegraphics[width=\linewidth]{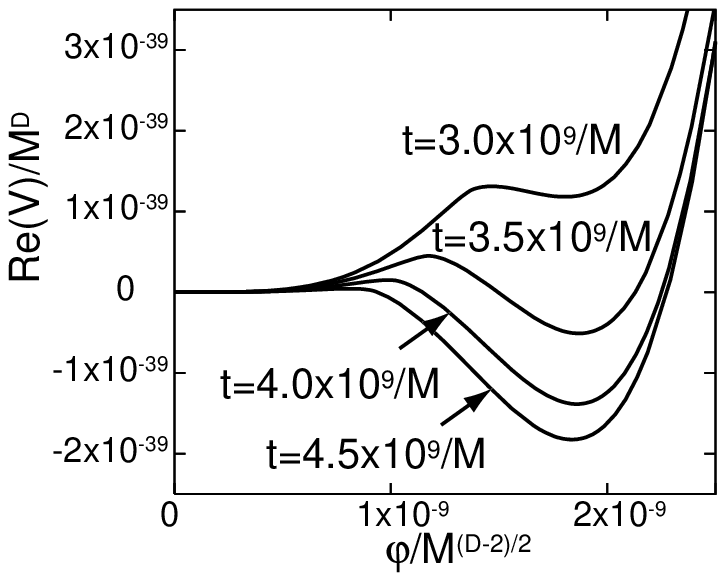}
\end{minipage}
\begin{minipage}{6.8cm}
\includegraphics[width=\linewidth]{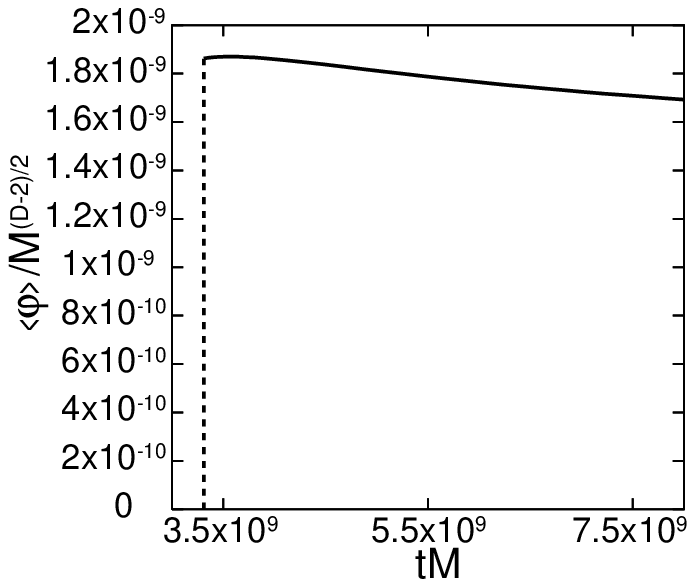}
\end{minipage}
\caption{\label{minimu0a0c1}Behaviour of the effective potential and the mass
gap for $\alpha=0$, $h_0=1$, $\xi=0$, $\mu_r=0$, $\lambda=0.1 M^{4-D}$ and
$D=3.8$.}
\end{center}
\end{figure}

\begin{figure}[t]
\begin{center}
\begin{minipage}{6.8cm}
\includegraphics[width=\linewidth]{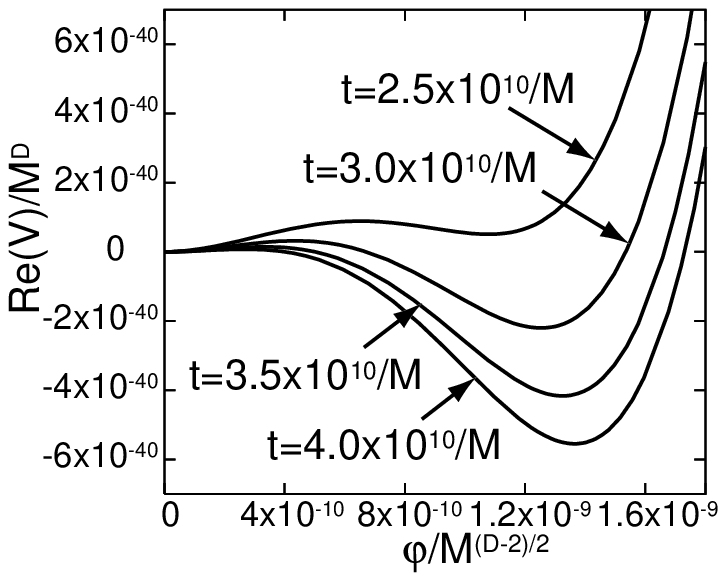}
\end{minipage}
\begin{minipage}{6.8cm}
\includegraphics[width=\linewidth]{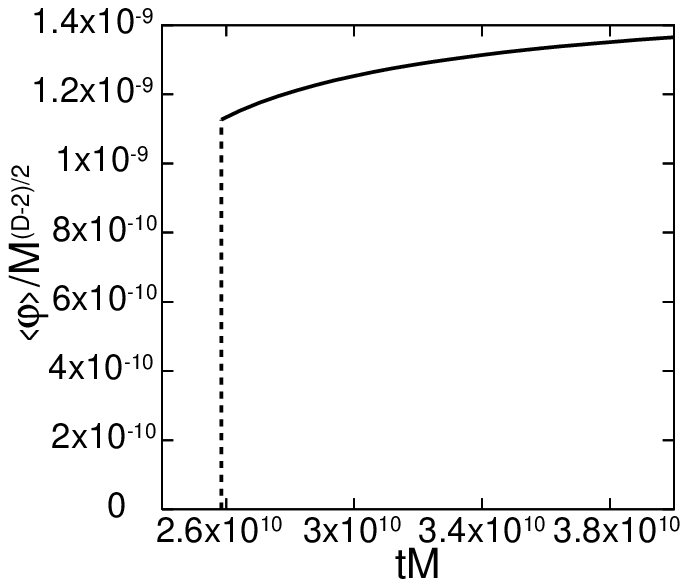}
\end{minipage}
\caption{\label{confmu0a0c0}Behaviour of the effective potential and the mass
gap for $\alpha=0$, $h_0=1$, $\xi=\xi_{\mbox{\scriptsize conformal}}$,
$\mu_r=0$, $\lambda=0.1 M^{4-D}$ and $D=3.8$.}
\end{center}
\end{figure}

Next we plot the behaviour of the effective potential for $\mu_r=0$ in Figs.
4, 5, 6 and 7. Here we consider the minimal and conformal gravitational
coupling,
$\xi=0$ and $\xi=\xi_{\mbox{\scriptsize conformal}}=(D-2)/[4(D-1)]$
respectively. (Of course, the scalar sector is conformally invariant
for correspondent choice of $\xi$ only when $\alpha=0$.)
For $\mu_r=0$ spontaneous symmetry breaking can not take place on the
classical level of the ordinary $\phi^4$ theory.
The radiative correction has an essential role to break the symmetry in
the ordinary theory, i.e. $\alpha=0$. It is known as a radiative symmetry
breaking.
First order phase transition is observed in Figs. 4 and 5.
However, the radiative correction has only a small effect. The expectation
value generated by the radiative symmetry breaking is extremely small,
as is shown in Figs. 4 and 5 at $\alpha=0$. A long time is necessary
until phase transition takes place.

\begin{figure}[t]
\begin{center}
\begin{minipage}{6.8cm}
\includegraphics[width=\linewidth]{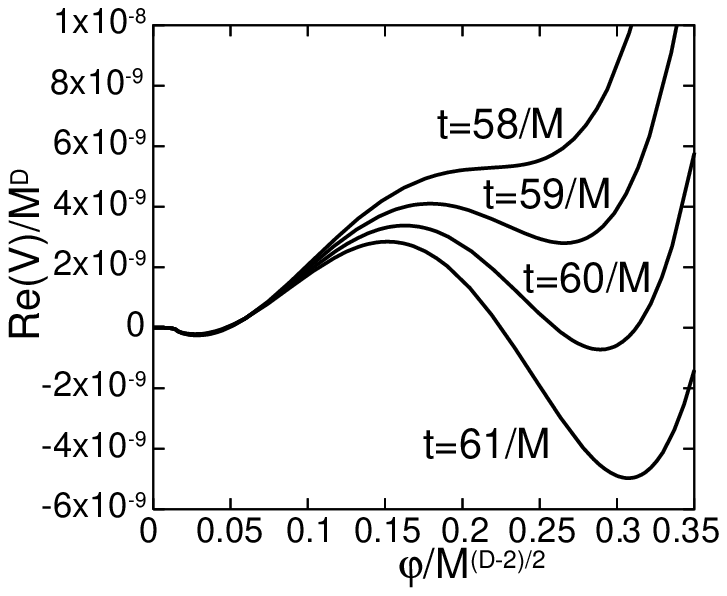}
\end{minipage}
\begin{minipage}{6.8cm}
\includegraphics[width=\linewidth]{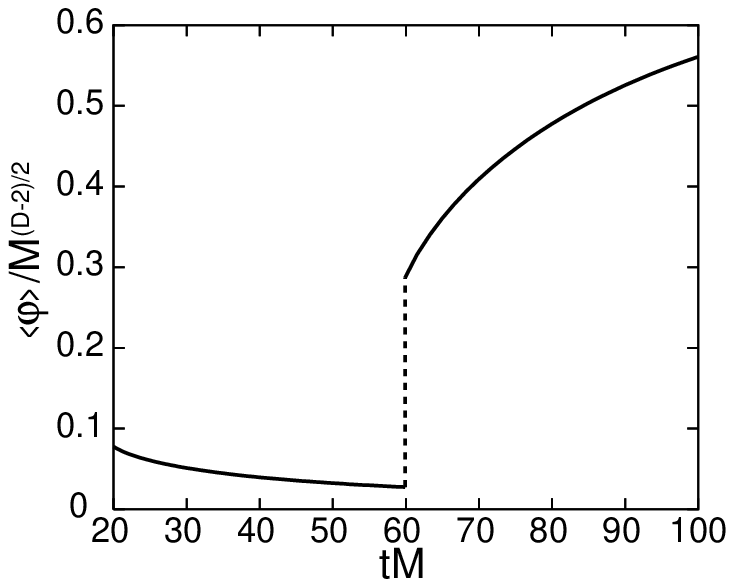}
\end{minipage}
\caption{\label{minimu0a1c1}Behaviour of the effective potential and the mass
gap for $\alpha=1$, $h_0=1$, $\xi=0$, $\mu_r=0$, $\lambda=0.1 M^{4-D}$ and
$D=3.8$.}
\end{center}
\end{figure}

\begin{figure}[t]
\begin{center}
\begin{minipage}{6.8cm}
\includegraphics[width=\linewidth]{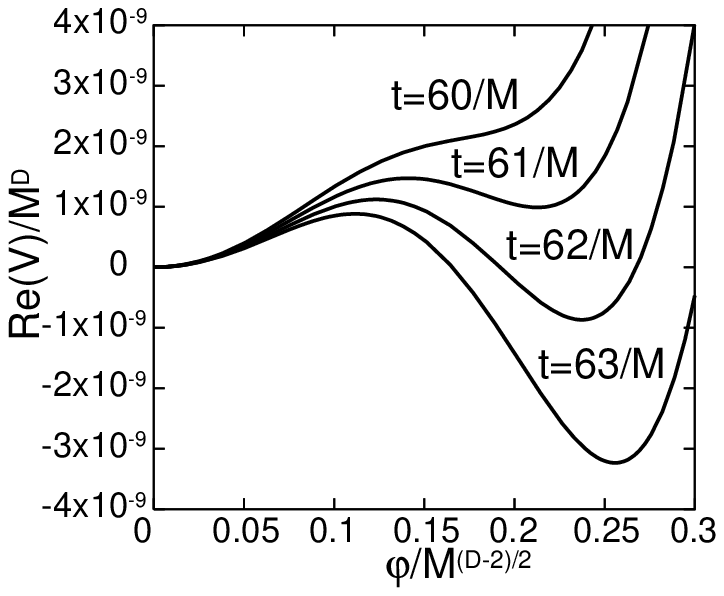}
\end{minipage}
\begin{minipage}{6.8cm}
\includegraphics[width=\linewidth]{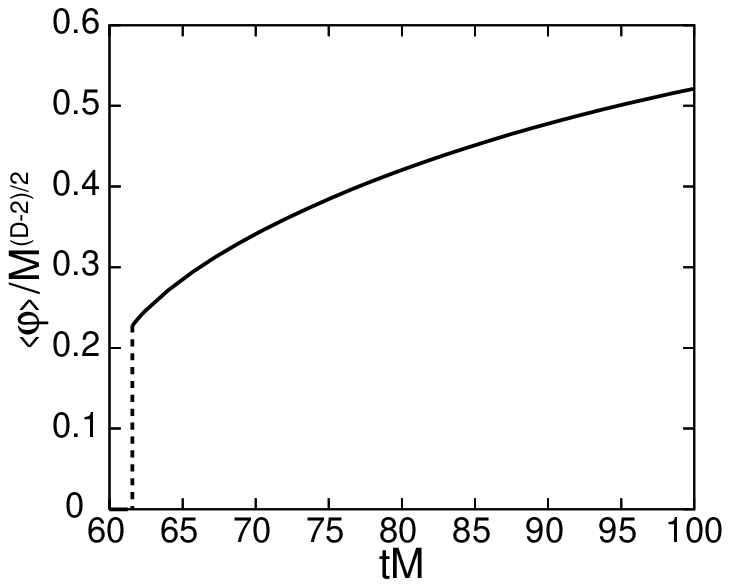}
\end{minipage}
\caption{\label{confmu0a1c1}Behaviour of the effective potential and the mass
gap for $\alpha=1$, $h_0=1$, $\xi=\xi_{\mbox{\scriptsize conformal}}$,
$\mu_r=0$, $\lambda=0.1 M^{4-D}$ and $D=3.8$.}
\end{center}
\end{figure}

In the case of a positive $\alpha$ the ratio of the radiative correction
to the classical part of the $\phi^4$ Lagrangian is enhanced as the
curvature decreases.
In Figs. 6 and 7 we also observe the first order phase transition and find
that the expectation value $\langle\phi\rangle$ is extremely enhanced
in comparison with the case $\alpha=0$.
For $\alpha=-1$ only a symmetric phase is realized at $\mu_r=0$ and then
radiative symmetry breaking does not occur.

As is shown in Figs. 6 and 7, the expectation value of $\phi$ has
non-negligible time dependence for the case $\mu=0$.
Hence we next consider the spacialy homogeneous but time dependent $\phi$ at
$\mu=0$ and $\alpha =1$.
Then we should solve the gap equation (\ref{Gap:Action}).
From Eq. (\ref{Gap:Action}) the equation of motion for $\phi$ is
obtained
\begin{eqnarray}
    && \frac{1}{\sqrt{-g}}
    \frac{\delta}{\delta \phi} \int d^D x \sqrt{-g} {\cal L}_{eff} =
    -\frac{\partial V(\phi)}{\partial \phi}
    - \frac{1}{\sqrt{-g}} \partial_{\mu} \sqrt{-g}
      \left(\frac{R}{M^2} \right)^{\alpha} \partial^{\mu} \phi
\nonumber \\
    && =  -\frac{\partial V(\phi)}{\partial \phi}
        -\left(\frac{R}{M^2} \right)^{\alpha} \left[
         \ddot{\phi}+(D-1)\frac{\dot{a}}{a}\dot{\phi}
         -\frac{2\alpha}{t}\dot{\phi}
        \right]
     =  0 .
\label{eqm}
\end{eqnarray}
To find an exact solution one needs to solve this equation for a general
time-dependent form of $\phi$. However, it is instructive to consider the
solution of Eq.(\ref{eqm}) for a special form of $\phi$.
In the present paper it is assumed that
\begin{equation}
      \phi(t)=\langle\phi(t)\rangle = v t^x,
      \ \ \phi^{;\mu}\phi_{;\mu} = \frac{x^2}{t^2} \langle \phi(t) \rangle^2 ,
\label{ass2}
\end{equation}
where $v$ is a constant parameter. Here we fix the parameter
$x$ at $1/2$ or $1$ and numerically solve the equation of motion for $\phi$.

Under the assumption (\ref{ass2}) Eq.(\ref{eqm}) reads
\begin{equation}
    \left[(x-1)+(D-1)h_0-2\alpha\right] \frac{x}{t^2}\phi(t)
        +\left(\frac{R}{M^2} \right)^{-\alpha}
        \frac{\partial V(\phi)}{\partial \phi} = 0 .
\label{eqm2}
\end{equation}
Thus the expectation value $\langle\phi\rangle$ is found by
observing the stationary point of the following function
\begin{equation}
    F[\phi]\equiv\left[(x-1)+(D-1)h_0-2\alpha\right] \frac{x}{2 t^2}\phi(t)^2
        +\left(\frac{R}{M^2} \right)^{-\alpha} V(\phi) .
\label{eff2}
\end{equation}
The solution is shown in Figs. 8 and 9.
As is seen in Figs. 8 and 9, we observe the first order phase transition
again.
In the case of the minimal gravitational coupling we find the two steps
of the first order phase transition for $x=1$. In Fig. 8 (a) the solution
becomes unstable for $tM<15.5$. The weak curvature approximation and/or
the assumption $\langle\phi(t)\rangle = vt$ can not be applicable for
such a small $t$.
In all the cases the mass scale $v$ depends on the time $t$. However, we
observe that the mass scale $v$ is almost static between $tM=80$ and $tM=90$
for $x=1$ and between $tM=90$ and $tM=100$ for $x=1/2$. It is expected
that there is a solution with gradually decreasing $x$ after the first
order phase transition near the time $tM=60$.
On the other hand, we observe in Figs. 6, 7, 8 and 9 that the time of the 
first order transition near $tM=60$ has no large dependence on $x$ in the 
present analysis at $\alpha=1$ and $h_0=1$. It seems to be a characteristic
feature of the theory with $\alpha=1$ in the FRW spacetime.

\begin{figure}[t]
\begin{center}
\begin{minipage}{6.8cm}
\begin{center}
\includegraphics[width=\linewidth]{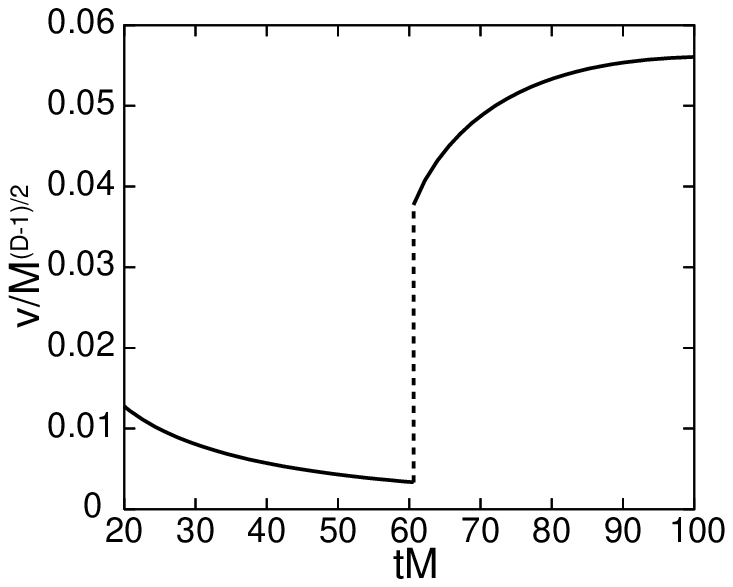}
\\
(a) $\langle \phi \rangle = v \sqrt{t},\  (x=1/2)$
\end{center}
\end{minipage}
\begin{minipage}{6.8cm}
\begin{center}
\includegraphics[width=\linewidth]{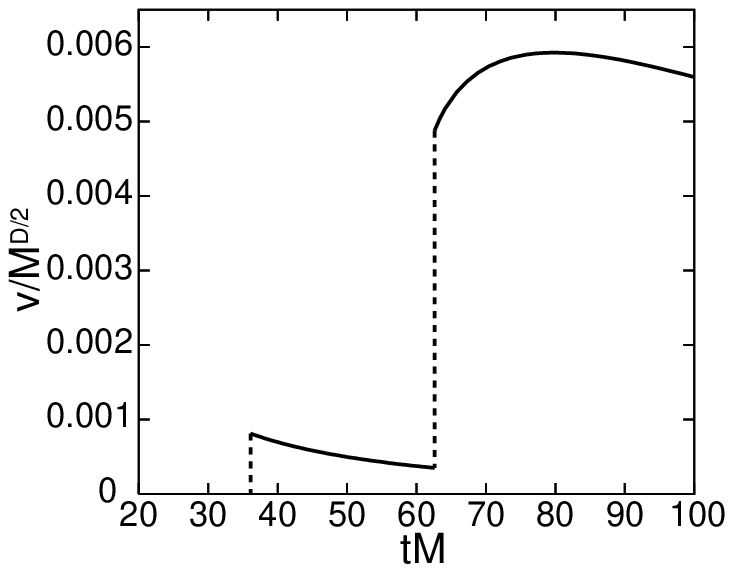}
\\
(b) $\langle \phi \rangle = v t,\  (x=1)$
\end{center}
\end{minipage}
\caption{\label{minimu0a1c1d1}Behaviour of the mass scale $v$ for $\alpha=1$,
$h_0=1$, $\xi=0$, $\mu_r=0$, $\lambda=0.1 M^{4-D}$ and $D=3.8$.}
\end{center}
\end{figure}

\begin{figure}[t]
\begin{center}
\begin{minipage}{6.8cm}
\begin{center}
\includegraphics[width=\linewidth]{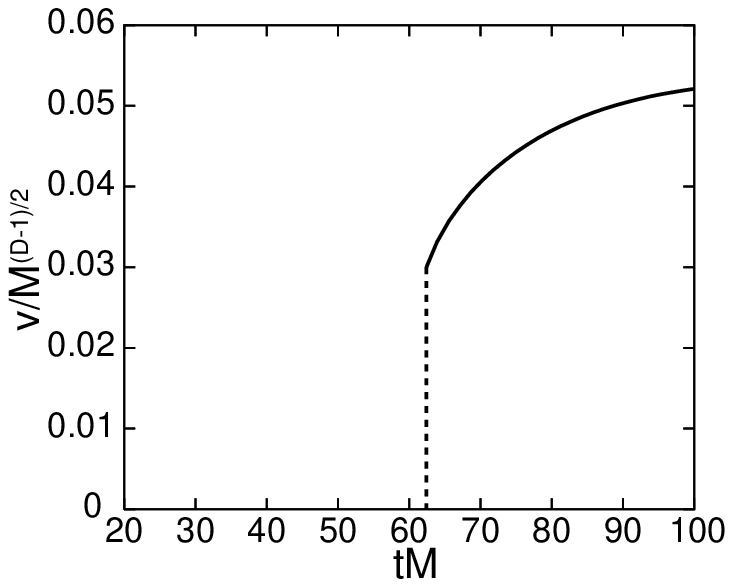}
\\
(a) $\langle \phi \rangle = v \sqrt{t},\  (x=1/2)$
\end{center}
\end{minipage}
\begin{minipage}{6.8cm}
\begin{center}
\includegraphics[width=\linewidth]{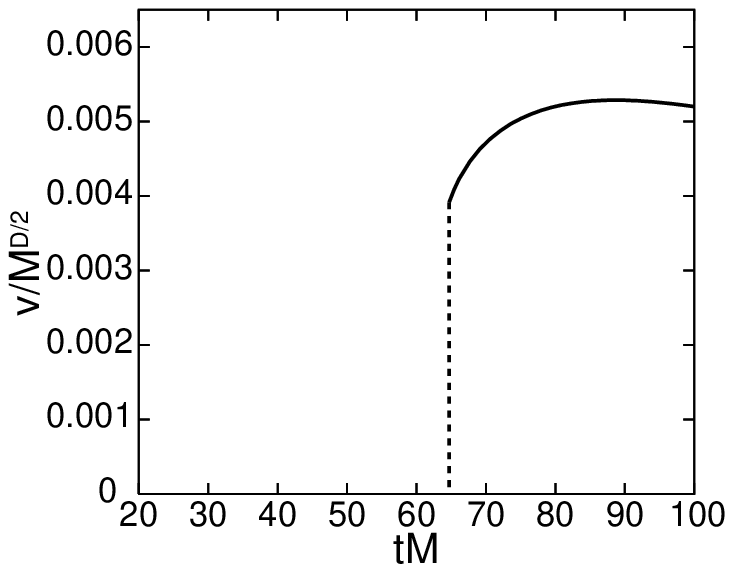}
\\
(b) $\langle \phi \rangle = v t,\  (x=1)$
\end{center}
\end{minipage}
\caption{\label{confmu0a1c1d1}Behaviour of the mass scale $v$ for $\alpha=1$,
$h_0=1$, $\xi=\xi_{\mbox{\scriptsize conformal}}$, $\mu_r=0$, $\lambda=0.1
M^{4-D}$ and $D=3.8$.}
\end{center}
\end{figure}

Hence, in the present section we numerically estimated phase structure
for scalar theory in curved spacetime under consideration.
The phase structure strongly depends on the sign of $\alpha$.
Compared with the usual $\phi^4$ theory in weakly curved space,
i.e. $\alpha=0$,
a positive $\alpha$ significantly enhances the symmetry breaking.
For a negative $\alpha$ the radiative symmetry breaking does not take
place.
As it was shown, mainly first-order curvature induced phase transitions may
occur here. 

In all figures we put $D=3.8$ and $h_0=1$. The spacetime curvature
is given by Eq.(\ref{r:frw}) and found to be $R\sim 5.0/t^2$.
The typical scale found to be $tM\sim 50$, 
$\langle\phi\rangle \sim 0.5 M^{(D-2)/2}$ 
in Figs. 1$\sim$3, 6$\sim$9 and $tM\sim 10^{10}$, 
$\langle\phi\rangle \sim 10^{-9}M^{(D-2)/2}$ in Figs. 4, 5.
The curvature scales for such time scales are given by 
$R\sim 0.002 M^2$ and $10^{-20}M^2$ respectively.
If we consider the electroweak scale Higgs boson, 
$\langle\phi\rangle \sim (100$GeV$)^{(D-2)/2}$, we obtain
the typical mass scale $M\sim 200$GeV in the former case and  
$M\sim 10^{11}$GeV in the latter case.
The cut-off scales which corresponds to $D=3.8$ are
$\Lambda\sim 1$TeV, near the scale of a large extra dimension, 
and $10^{12}$GeV, respectively. 
For both cases the typical scale of time and curvature are given by
$t\sim 0.1($GeV${}^{-1})$ and $R\sim 100($GeV${}^2)$.
Thus the quantum correction induced by the curvature effect has an 
interesting contribution to the effective action at the very early 
stage of the universe.
The time scale of the phase transition depends on the coupling 
constant $\lambda$. For a stronger coupling the radiative correction
is enhanced and a smaller time scale is obtained.

Thus we consider the regularized theory and calculate the effective
potential in a D-dimensional spacetime with $D=3.8$.
As is shown in the previous section, we obtain the finite effective 
potential (\ref{alrd4}) in four spacetime dimensions. Here we show 
some results in four dimensions. Evaluating the effective potential 
(\ref{alrd4}) we draw the solution of the gap equation. The first 
order phase transitions is observed as is shown in Fig. 10. 
Compared with the case $D=3.8$ (Figs. 6 and 7) the critical time 
becomes late and the the two steps transition does not realised
in four dimensions.

\begin{figure}[t]
\begin{center}
\begin{minipage}{6.8cm}
\begin{center}
\includegraphics[width=\linewidth]{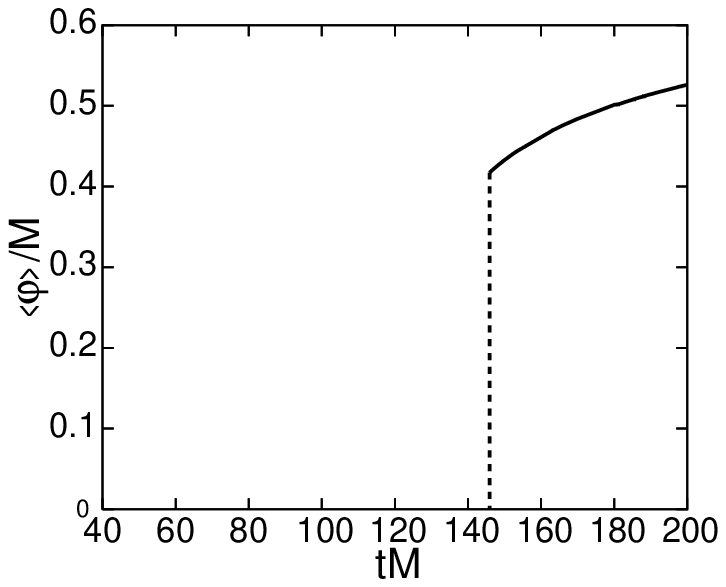}
\\
(a) $\xi=0$
\end{center}
\end{minipage}
\begin{minipage}{6.8cm}
\begin{center}
\includegraphics[width=\linewidth]{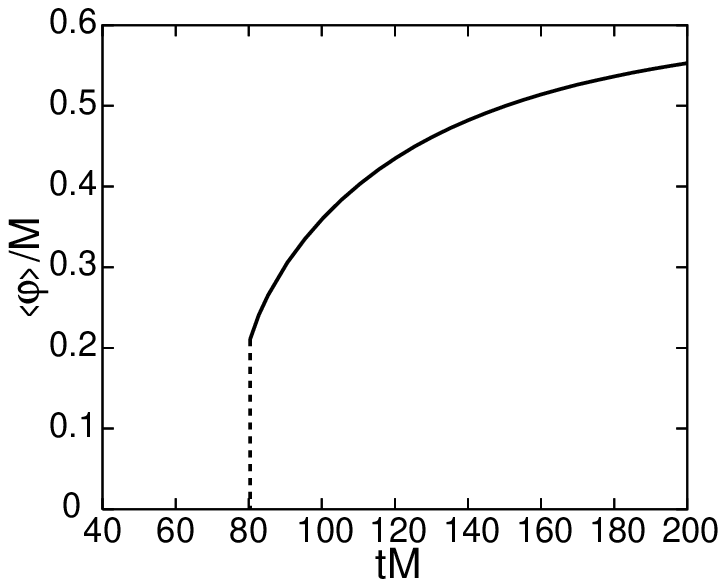}
\\
(b) $\xi=\xi_{\mbox{\scriptsize conformal}}$
\end{center}
\end{minipage}
\caption{\label{gapd4}Behaviour of the effective potential and the mass
gap for $\alpha=1$, $h_0=1$, $\mu_r=0$, $\lambda=0.1$ in $D=4$.}
\end{center}
\end{figure}

The main lesson of this consideration is that it is unlikely that
radiative corrections may strongly influence the dark energy universe in such
theory. However, quantum effects may become quite important
for phantom universe where Big Rip may occur \cite{bigrip, bigrip2}. 
It was already
demonstrated that in such circumstances already quantum effects of free fields
may prevent Big Rip occurrence \cite{elizalde, elizalde2, elizalde3, elizalde4}.
However, in order to analyse this problem in our framework one needs the 
calculation of the one-loop effective action at strong curvature.

\section{Dynamical mechanism to solve the  cosmological constant problem}

\subsection{Dynamical cosmological constant theory: exact example}

There are some proposals to solve the cosmological constant problem dynamically
\cite{dolgov}.
One of the possible solutions of the cosmological constant
problem is pointed out by  Mukohyama and  Randall \cite{MR}.
In \cite{MR}, the following type of the action similar to the one under
consideration has been proposed:
\begin{equation}
\label{MR1}
I=\int d^4 x\sqrt{-g}\left[\frac{R}{2\kappa^2} + \alpha_0 R^2 +
\frac{\left(\kappa^4\partial_\mu \varphi
\partial^\mu \varphi\right)^q}{2q \kappa^4 f(R)^{2q-1}} - V(\varphi)\right]\ ,
\end{equation}
where $f(R)$ is a proper function of the scalar curvature $R$. When the
curvature is small, it is assumed
$f(R)$ behaves as
\begin{equation}
\label{MRIO1}
f(R)\sim \left(\kappa^2 R^2 \right)^m\ .
\end{equation}
Here  $m$ is positive.
When the curvature is small, the vacuum energy, and therefore the value of the
potential becomes small.
Then one may assume, for the small curvature, $V(\varphi)$ behaves as
\begin{equation}
\label{MRIO2}
V(\varphi)\sim V_0 \left(\varphi - \varphi_c\right)\ .
\end{equation}
Here $V_0$ and $\varphi_c$ are constants. If $q>1/2$, the factor in front of
the kinetic term of
$\varphi$ in (\ref{MR1}) becomes large. This makes the time development of the
scalar field $\varphi$
very slow and it is expected that $\varphi$ does not reach $\varphi_c$. This
model may explain the
acceleration of the present universe. The model  (\ref{MR1}) is also expected
to be stable
under the radiative corrections. In fact, when the curvature is small and the
time development of the
curvature can be neglected, if we rescale the scalar field $\varphi$ as
$\varphi\to R^{m(2q-1)/q}\varphi$, the curvature in the kinetic term can be
absorbed into the
redefinition of $\varphi$ and there appear factors including $R^{m(2q-1)/q}$.
Therefore
if $m(2q-1)/q>0$, the interactions could be suppressed when the curvature is
small and there will not
appear the radiative correction to the vacuum energy except the one loop
corrections.

We now show that there is an exactly solvable model, which realizes the
scenario in \cite{MR}.
Assume $m=q=1$ and (\ref{MRIO1}) and (\ref{MRIO2}) are exact:
\begin{equation}
\label{MRIO3}
f(R)= \beta R^2 \ ,\quad
V(\varphi)= V_0 \left(\varphi - \varphi_c\right)\ .
\end{equation}
Here $\beta$ is a constant.  $R^2$ term is neglected by putting $\alpha_0=0$ in
(\ref{MR1}) since
   the curvature is small.
Then by the variation over the metric, one obtains
\begin{eqnarray}
\label{MRIO4}
0&=& -\frac{1}{2}\left\{\frac{R}{2\kappa^2} + \frac{\partial_\rho\varphi
\partial^\rho\varphi}{2\beta R^2}
   - V_0 \left(\varphi - \varphi_c\right)\right\}g_{\mu\nu}
   + \frac{R_{\mu\nu}}{2\kappa^2} - \frac{\partial_\rho \varphi \partial^\rho
\varphi R_{\mu\nu}}{\beta R^3} \nonumber \\
&& + \frac{\partial_\mu \varphi \partial_\nu \varphi}{2\beta R^2}
   - \left(\nabla_\mu \nabla_\nu - g_{\mu\nu} \nabla^2\right)\left(
\frac{\partial_\rho\varphi \partial^\rho\varphi}{\beta R^3}\right)\ .
\end{eqnarray}
On the other hand,  the variation over $\varphi$ gives
\begin{equation}
\label{MRIO6}
0=-\partial_\mu \left(\sqrt{-g}\frac{\partial^\mu\varphi}{\beta R^2}\right) -
V_0 \sqrt{-g}\ .
\end{equation}
We assume the four dimensional FRW metric with flat spatial part:
\begin{equation}
\label{MR11}
ds^2 = dt^2 - a(t)^2 \sum_{i=1,2,3} \left(dx^i\right)^2\ .
\end{equation}
It is chosen that $\varphi$ only depends on the time coordinate $t$.
Then the $(t,t)$-component of (\ref{MRIO4}) gives
\begin{eqnarray}
\label{MRIO7}
0&=& - \frac{3H^2}{2\kappa^2} + \frac{{\dot \varphi}^2}{144\beta \left(\dot H +
H^2\right)^2}
+ V_0\left(\varphi - \varphi_c\right) \nonumber \\
&& + H\frac{d}{dt}\left(\frac{{\dot \varphi}^2}{12\beta \left(\dot H +
H^2\right)^2}\right)
   - \frac{{\dot \varphi}^2\left(\dot H + H^2\right)}{72\beta \left(\dot H +
H^2\right)^3}\ ,
\end{eqnarray}
where $H$ is the Hubble rate.
On the other hand, (\ref{MRIO6}) gives
\begin{equation}
\label{MRIO8}
0=-\frac{d}{dt}\left(\frac{a^3 \dot \varphi}{36\beta \left(\dot H +
H^2\right)^2}\right)
   - V_0 a^3\ .
\end{equation}
A solution of Eqs.(\ref{MRIO7}) and (\ref{MRIO8}) is given by
\begin{equation}
\label{MRIO9}
a=a_0 t^{h_0}\ \left(H=\frac{h_0}{t}\right)\ ,\quad \varphi=\varphi_c +
\frac{\varphi_0}{t^2}\ ,
\end{equation}
when $h_0>0$. In case $h_0<0$, this scale factor $a=a_0 t^{h_0}$ (\ref{MRIO9})
does not express expanding universe but shrinking one.
If  the direction of time is changed as $t\to -t$, the expanding universe
emerges with  scale factor
   $a=a_0(-t)^{h_0}$. In the expression, however, since $h_0$ is not always an
integer, $t$ should
be negative so that the scale factor should be real. To avoid this apparent
difficulty, we may further shift
the origin of the time as $t\to -t \to t_s - t$. Then the time $t$ can be
positive as long as $t<t_s$.
When $h_0<0$, instead of (\ref{MRIO9}), one may propose
\begin{equation}
\label{MRIO9b}
a=a_0 \left(t_s - t\right)^{h_0}\ \left(H=\frac{h_0}{t_s-t}\right)\ ,
\quad \varphi=\varphi_c + \frac{\varphi_0}{\left(t_s -t\right)^2}\ ,
\end{equation}
The assumption (\ref{MRIO9}) or (\ref{MRIO9b}) reduces Eqs.(\ref{MRIO7}) and
(\ref{MRIO8}) to the algebraic
equations:
\begin{eqnarray}
\label{MRIO10}
0&=& - \frac{3h_0^2}{2\kappa^2} + \frac{\varphi_0^2h_0}{36\beta\left( - h_0 +
2h_0^2\right)^3}
+ V_0\varphi_0 \ ,\\
\label{MRIO11}
0&=&\frac{\varphi_0\left(3h_0 + 1\right)}{18\beta \left(-h_0 + 2h_0^2\right)^2}
- V_0\ ,
\end{eqnarray}
which gives
\begin{eqnarray}
\label{MRIO12}
\varphi_0^2 &=& \frac{54\beta \left( -1 + 2h_0\right)^3
h_0^4}{\kappa^2\left(12h_0^2 - 2h_0 - 1\right)}\ ,\nonumber \\
V_0&=&\pm \frac{3h_0 + 1}{\sqrt{6\kappa^2\left(12h_0^2 - 2h_0 - 1\right)
\left(-1 + 2h_0\right)}}\ .
\end{eqnarray}
Since $\varphi_0^2$ should be positive, one finds
\begin{eqnarray}
\label{MRIO13}
& \mbox{when}\ \beta>0\ ,\quad
&\frac{1-\sqrt{13}}{12}<h_0<\frac{1+\sqrt{13}}{12}\
\mbox{or}\ h_0\geq \frac{1}{2}\ ,\nonumber \\
& \mbox{when}\ \beta<0\ ,\quad &h_0<\frac{1-\sqrt{13}}{12}\ \mbox{or}\
\frac{1+\sqrt{13}}{12}<h_0\leq \frac{1}{2}\ .
\end{eqnarray}
By properly choosing the parameter $\beta$ and $V_0$, we can always obtain a
solution as in
(\ref{MRIO9}) or (\ref{MRIO9b}).
For example, if \begin{equation}
\label{AA1}
h_0=-\frac{1}{60}\ ,
\end{equation}
which gives the following value of the parameter $w$ of the equation of state
\begin{equation}
\label{AA2}
w= -1 + \frac{2}{3h_0}=- 1.025\ ,
\end{equation}
we find
\begin{equation}
\label{AA3}
\kappa V_0=\pm \frac{19}{34}\sqrt{\frac{15}{31}}=\pm 0.388722...\ .
\end{equation}
It is interesting that the value of $w$  (\ref{AA2}) is consistent with the
observed one.

For $h_0>0$ case, corresponding to (\ref{MRIO9}), since $R=6\dot H + 12 H^2$,
the curvature $R$
decreases as $t^{-2}$ with time $t$. Eq.(\ref{MRIO9}) tells that $\varphi$
approaches to $\varphi_c$
but does not arrive at $\varphi_c$ in a finite time, as expected in \cite{MR}.

As seen from the expression of $H$ in (\ref{MRIO9}) or (\ref{MRIO9b}), if we
substitute
the value of the age of the present universe $10^{10}$years$\sim
(10^{-33}$eV$)^{-1}$ into $t$ or $t_s-t$,
the observed value of $H$ could be reproduced, which could explain the
smallness of the effective
cosmological constant $\Lambda\sim H^2$.
Note that even if there is no potential term, that is, $V_0=0$, when $\beta<0$,
there is a solution
\begin{equation}
\label{MRIO13b}
h_0=-\frac{1}{3}<\frac{1-\sqrt{13}}{12}=-0.2171...\ ,
\end{equation}
which gives the parameter $w$ of the equation of state: $w=-3$,
although  $w$ is not realistic.

\subsection{Cosmological constant problem in scalar theory non-linearly coupled
with curvature}
Here we consider the solution of the cosmological constant problem
in the theory (\ref{a0}).
As is shown in the previous section,
one could expect that the radiative correction is suppressed
when the curvature is small if $\alpha<0$. As in the case of (\ref{MR1}), if we
rescale $\phi$ by
$\phi\to (R/M^2)^{-\alpha/2}\phi$, the kinetic term  becomes standardly
normalized one and
does not include the curvature. The $\phi^4$ interaction terms, however,
   include
the factor $(R/M^2)^{-\alpha}$ and may be suppressed for the small curvature.
Hence,
there would not appear the radiative correction to the vacuum energy
except the one-loop corrections.

As an example of the solvable case, we put $\alpha=-1$ and $\mu_0^2=0$.
The variation over the metric gives
\begin{eqnarray}
\label{MRIO15}
0&=&-\frac{1}{2}\left[ \frac{R}{2\kappa^2} - \frac{M^2}{R}\left\{
\frac{1}{2}\partial_\rho \phi \partial^\rho \phi - \frac{\xi R}{2} \phi^2
   + \frac{\lambda}{4!}\phi^4\right\}\right]g_{\mu\nu}
   + \frac{R_{\mu\nu}}{2\kappa^2} \nonumber \\
&& + \frac{M^2}{2R}\partial_\mu \phi \partial_\nu \phi
   - \frac{M^2 R_{\mu\nu}}{R^2}\left\{
    \frac{1}{2}\partial_\rho \phi \partial^\rho \phi +
\frac{\lambda}{4!}\phi^4\right\} \nonumber \\
&& - \left(\nabla_\mu \nabla_\nu - g_{\mu\nu}
\nabla^2\right)\left\{\frac{M^2}{R^2}\left(
\frac{1}{2}\partial_\rho \phi \partial^\rho \phi +
\frac{\lambda}{4!}\phi^4\right)\right\}\ .
\end{eqnarray}
Scalar field equation is
\begin{equation}
\label{MRIO16}
0=- \frac{1}{\sqrt{-g}}\partial_\mu\left(\sqrt{-g}\frac{\partial^\mu
\phi}{R}\right)
   + \frac{\lambda}{3!R}\phi^3 + \xi\phi \ .
\end{equation}
By assuming the FRW metric  (\ref{MR11}) and $\phi=\phi(t)$, Eqs.(\ref{MRIO15})
and (\ref{MRIO16})
have the following forms:
\begin{eqnarray}
\label{MRIO17}
0&=& -\frac{3H^2}{2\kappa^2} + \frac{M^2\left(\frac{1}{2}{\dot \phi}^2 -
\frac{\lambda}{4!}\phi^4\right)}
{12\left(\dot H + 2H^2\right)} - \frac{M^2\left(\dot H + H^2\right)
\left(\frac{1}{2}{\dot \phi}^2 + \frac{\lambda}{4!}\phi^4\right)}
{12\left(\dot H + 2H^2\right)^2} \nonumber \\
&& + \frac{M^2 \xi}{4}\phi^2 - \frac{M^2}{12}\frac{d}{dt}\left(
\frac{\left(\frac{1}{2}{\dot \phi}^2 + \frac{\lambda}{4!}\phi^4\right)}
{\left(\dot H + 2H^2\right)^2}\right)\ , \\
\label{MRIO18}
0&=& - \frac{d}{dt}\left(\frac{a^3 \dot \phi}{\dot H + 2H^2} \right)
   + \frac{\lambda a^3 \phi^3}{3!\left(\dot H + 2H^2\right)} + \xi \phi a^3\ .
\end{eqnarray}
With the Anzats
\begin{equation}
\label{MRIO19}
H=\frac{h_0}{t}\ ,\quad \phi=\frac{\phi_0}{t}\ ,
\end{equation}
when $h_0>0$ or
\begin{equation}
\label{MRIO19b}
H=\frac{h_0}{t_s - t}\ ,\quad \phi=\frac{\phi_0}{t_s - t}\ ,
\end{equation}
when $h_0<0$.
Eqs.(\ref{MRIO17}) and (\ref{MRIO18}) reduce to the algebraic equations:
\begin{eqnarray}
\label{MRIO20}
0&=& - \frac{3h_0^2}{2\kappa^2} + \frac{M^2}{12\left( - h_0 + 2h_0^2\right)}
\left(\frac{1}{2}\phi_0^2 - \frac{\lambda}{4!}\phi_0^4\right) \nonumber \\
&& - \frac{M^2\left(-h_0 + h_0^2\right)}{12\left( - h_0 + 2h_0^2\right)^2}
\left(\frac{1}{2}\phi_0^2 + \frac{\lambda}{4!}\phi_0^4\right)
+ \frac{M^2\xi}{4}\phi_0^2 \ ,\\
\label{MRIO21}
0&=& \frac{h_0\phi_0}{2\left(-h_0 + 2h_0^2\right)} + \frac{\lambda
\phi_0^3}{36\left(-h_0 + h_0^2\right)}
   + \xi\phi_0\ ,
\end{eqnarray}
which give
\begin{eqnarray}
\label{MRIO22}
\phi_0^2&=&\frac{3}{\displaystyle\kappa^2\left\{\frac{8-9h_0}{24\left(-h_0 +
h_0^2\right)^2}
- \frac{\left(4-7h_0\right)\xi}{\left( -h_0 + 2h_0^2\right)h_0}\right\}}\ ,\\
\label{MRIO23}
\lambda&=& -6\kappa^2h_0 \left\{ 1 - 2\left(1-2h_0\right)\xi\right\} \nonumber
\\
&& \times \left\{\frac{8-9h_0}{24\left(-h_0 + h_0^2\right)^2}
- \frac{\left(4-7h_0\right)\xi}{\left( -h_0 + 2h_0^2\right)h_0}\right\}\ .
\end{eqnarray}
Therefore, with the proper choice of parameters, the solution in the form of
(\ref{MRIO19}) or (\ref{MRIO19b}) follows.
We should note the behavior of $H$  (\ref{MRIO19}) or (\ref{MRIO19b}) is almost
identical with
that in (\ref{MRIO9}) or (\ref{MRIO9b}).

One may consider the model including two scalar fields $\varphi$ and $\phi$ as
\begin{eqnarray}
\label{MRIO23b}
I&=&\int d^4 x\sqrt{-g}\left[\frac{R}{2\kappa^2}  + \frac{\partial_\mu \varphi
\partial^\mu \varphi}
{2\beta R^2} - V_0\left(\varphi - \varphi_c\right) \right. \nonumber \\
&& \left. - \frac{M^2}{R}\left\{
\frac{1}{2}\partial_\mu \phi \partial^\mu \phi + \frac{\xi R}{2} \phi^2
   + \frac{\lambda}{4!}\phi^4\right\}\right]\ .
\end{eqnarray}
Even in this case, for FRW metric (\ref{MR11}) and assuming (\ref{MRIO9}) or
(\ref{MRIO9b})
and/or (\ref{MRIO19}) or (\ref{MRIO19b}), we obtain (\ref{MRIO18}) and
(\ref{MRIO21}).
The equation corresponding to (\ref{MRIO17}) and/or (\ref{MRIO20}) has the
following form:
\begin{eqnarray}
\label{MRIO24}
0&=& -\frac{3H^2}{2\kappa^2} + \frac{M^2\left(\frac{1}{2}{\dot \phi}^2 -
\frac{\lambda}{4!}\phi^4\right)}
{12\left(\dot H + 2H^2\right)} - \frac{M^2\left(\dot H + H^2\right)
\left(\frac{1}{2}{\dot \phi}^2 + \frac{\lambda}{4!}\phi^4\right)}
{12\left(\dot H + 2H^2\right)^2} \nonumber \\
&& + \frac{M^2 \xi}{4}\phi^2 - \frac{M^2}{12}\frac{d}{dt}\left(
\frac{\left(\frac{1}{2}{\dot \phi}^2 + \frac{\lambda}{4!}\phi^4\right)}
{\left(\dot H + 2H^2\right)^2}\right) \nonumber \\
&& + \frac{M^2}{12\left( - h_0 + 2h_0^2\right)}
\left(\frac{1}{2}\phi_0^2 - \frac{\lambda}{4!}\phi_0^4\right) \nonumber \\
&& - \frac{M^2\left(-h_0 + h_0^2\right)}{12\left( - h_0 + 2h_0^2\right)^2}
\left(\frac{1}{2}\phi_0^2 + \frac{\lambda}{4!}\phi_0^4\right)
+ \frac{M^2\xi}{4}\phi_0^2 \ .
\end{eqnarray}
Then by properly choosing the parameters, we may obtain an exact solution for
cosmological constant again.

Hence, we demonstrated that the mechanism proposed in ref.\cite{MR} to
dynamically solve the cosmological constant problem is naturally realized also
for the class of models investigated in this work.

The final remark is in order.
In \cite{Dolgov} (see also \cite{Ford}), the cosmological constant damping
in the model contained $\phi^2 R$ term in the Lagrangian
has been investigated. It has been shown the possibility to solve the 
cosmological constant problem. Although the problem could
be solved in the model, the effective Newton constant $G_{\rm eff}$
suffers the correction
like $G_{\rm eff}\sim G_0 /\left(1 + a \phi^2\right)$ with a constant $a$.
Here $G_0$ is the bare Newton constant.
In the solution of model  \cite{Dolgov}, $\phi$ behaves as a linear function
of the time $t$.
Naturally $t$ could be the age of the present universe,
$t\sim \left(10^{-33}\,\mbox{eV}\right)^{-1}$,
the correction becomes very large and the effective Newton constant
becomes unnaturally small.
In the model under consideration (\ref{a0}) with (\ref{I2}), similar term 
is included but due to the factor
$R^\alpha$, if $\alpha>0$, the correction is suppressed since
$R\sim \left(10^{-33}\,\mbox{eV}\right)^2$ and $R$ is very small.
In general $\alpha$ can be also 
negative. In the solvable case in (\ref{MRIO15}), the correction could be
$R^{-2}\left\{\frac{1}{2}\partial_\rho \phi \partial^\rho \phi
+ \frac{\lambda}{4!}\phi^4\right\}$. Since the $R\sim t^{-2}$
or $\left(t_s -t\right)^{-2}$,
by combining (\ref{MRIO19}) or (\ref{MRIO19b}),
the correction takes the order of unity.
Then the effective Newton constant could receive the correction but
it is finite, which does not lead to the instabilities of the above sort 
(smallness of gravitational constant).

\section{Conclusion}

In summary, we studied the scalar self-interacting theory whose Lagrangian
non-linearly couples with some power of the curvature as the effective field
theory. The one-loop effective action at weak curvature is found here using
specific regularization scheme. The phase structure of the theory is carefully
investigated, the examples of curvature-induced phase transitions
are presented numerically (for the case of positive power of curvature). The
comparison with the ordinary scalar self-interacting theory in curved space is
done. For negative value of
$\alpha$ the radiative symmetry breaking does not take place.
It remains to develop the same one-loop formulation at strong curvature
as in such a case the phase transitions maybe considered in the vicinity
of Big Rip.

The dynamical mechanism to explain the origin of the extremely small
cosmological constant  in the spirit of papers \cite{dolgov, MR} is presented.
It is shown that in the class of the models under consideration (including the
case with two scalars) such a proposal
maybe realized quite successfully. Unfortunately, despite the preliminary
expectations
the radiative corrections for the theory under consideration are not relevant
in the study of the dynamical cosmological constant.

There are many directions where our approach maybe generalised. In particularly,
it is known that phase structure of NJL model in curved spacetime is quite rich
(for a review, see \cite{IMO}). Hence, it would be extremely interesting to
consider the NJL-like models non-linearly coupled with power of the curvature
in the same spirit. Such models maybe
even more attractive due to fact that $1/N$ expansion is more reliable
than one-loop approximation in non-renormalizable theories. One can also
consider Yang-Mills theory (or Born-Infeld theory) coupled to curvature powers
in the same way. Definitely, this may bring the number of the interesting
applications in such issues as dark energy and cosmological constant.

From another point, the new matter-gravity coupling \cite{NO} maybe also
considered as kind of modification of gravitation itself. In such a situation,
when theory somehow deviates from General Relativity,
it is not clear apriori if the metric formalism and first-order formulation
should lead to the same physical picture \cite{FFV}. (Indeed,
unlike to General Relativity where metric and Palatini formulations basically
coincide, it is not the case for  our theory (1)\cite{NO}
already on classical level \cite{mauro}). It would be extremely interesting to
develop the one-loop effective action formulation
in the same way as the one developed in this paper when an external gravity is
formulated in Palatini form. This will be done elsewhere.

\ack

This research has been  supported in part by JSPS Grant-in-Aid
for Scientific Research, No.~13135208 (S.N.) and in part
 by RFBR grant 03-01-00105 and  LRSS
grant 1252.2003.2 (S.D.O.)

\end{document}